\documentclass[a4paper,12pt]{iopart}
\usepackage{color} 
\usepackage[T1]{fontenc}
\usepackage[utf8]{inputenc}
\usepackage{graphicx}
\usepackage{textcomp}
\usepackage{bm}
\usepackage{iopams}
\usepackage{dsfont}
\usepackage{mathrsfs}
\usepackage{fixmath}
\usepackage{exscale}
\usepackage[colorlinks=false, pdfborder={0 0 0}]{hyperref}
\newcommand{\eqref}[1]{(\ref{#1})}

\newcommand{\imag}{\rmi} 

\newcommand{\mrm}[1]{\mathrm{#1}}
\newcommand{\dd}{\mathrm{d}} 
\newcommand{\pdd}{\mathrm{\partial}} 

\newcommand{\pdiff}[3][]{\frac{\pdd^{#1} #2}{\pdd #3^{#1}}}

\newcommand{\qty}[1]{\left(#1\right)}
\newcommand{\sqty}[1]{\left[#1\right]}

\newcommand{\abs}[1]{\left|#1\right|}

\begin{document}
	\title[Diffractive focusing of a uniform Bose-Einstein condensate]{Diffractive focusing of a uniform Bose-Einstein condensate}
	\author{
		Patrick Boegel$\mathrm{^1}$, 
		Matthias Meister$\mathrm{^{2}}$, 
		Jan-Niclas Siemß$\mathrm{^{3,4}}$,\\
		Naceur Gaaloul$\mathrm{^4}$, 
		Maxim A. Efremov$\mathrm{^{1,2}}$, 
		and Wolfgang P. Schleich$\mathrm{^{1,2,5}}$}
	
	\address{
		$\mathrm{^1}$Institut f\"ur Quantenphysik and Center for Integrated Quantum Science and Technology (IQ$^{\rm ST}$), Universit\"at Ulm, D-89069 Ulm, Germany
	}
	\address{$\mathrm{^2}$Institute of Quantum Technologies, German Aerospace Center (DLR), D-89077 Ulm, Germany}
	\address{
		$\mathrm{^3}$Institut f\"ur Theoretische Physik, Leibniz Universit\"at Hannover, D-30167 Hannover, Germany
	}
	\address{$\mathrm{^4}$Institut f\"ur Quantenoptik, Leibniz Universit\"at Hannover, Welfengarten 1, D-30167 Hannover, Germany}
	\address{$\mathrm{^5}$Hagler Institute for Advanced Study at Texas A$\&$M University, Texas A$\&$M AgriLife Research, Institute for Quantum Science and Engineering (IQSE), and Department of Physics and Astronomy, Texas A$\&$M University, College Station, Texas 77843-4242, USA}

	\ead{patrick.boegel@uni-ulm.de, gaaloul@iqo.uni-hannover.de}

	\vspace{10pt}
	
	\begin{abstract}
		We propose a straightforward implementation of the phenomenon of diffractive focusing with uniform atomic Bose-Einstein condensates. Both, analytical as well as numerical methods not only illustrate the influence of the atom-atom interaction on the focusing factor and the focus time, but also allow us to derive the optimal conditions for observing focusing of this type in the case of interacting matter waves.
	\end{abstract}
	
	
	
	\vspace{2cm}
	Version: 13 August 2021
	
	\maketitle

	\section{Introduction}
	
	When focusing an electromagnetic wave, the position and the strength of the focus are typically controlled by a lens which imprints a position-dependent phase on the incoming wave. However, focusing is possible even without a lens, namely by employing the concept of diffractive focusing, where the focus is a consequence of the non-trivial shape of the initial wave function. In this article we extend this concept to non-linear matter waves and show how it can be experimentally realized with Bose-Einstein condensates (BECs).
	
	Already in 1816 A.~J.~Fresnel realized \cite{fresnel_diffractive_first, born_principles_1999} that light passing through a circular aperture creates a bright spot on the symmetry axis before it starts to spread. This type of focusing is nowadays known as diffractive focusing and was originally described for light by the two-dimensional (2D) paraxial Helmholtz equation \cite{guenther_modern_2015}.
	The same effect occurs for matter waves in one or multiple dimensions. Indeed, since the equations for electromagnetic fields within the paraxial approximation have a form similar to the Schr\"odinger equation of a free particle, diffractive focusing was studied
	~\cite{case_diffractive_2012, goncalves_single-slit_2017} and successfully observed for water waves and plasmons~\cite{weisman_diffractive_2017}, as well as for atoms~\cite{leavitt_single-slit_1969}, electrons~\cite{jonsson_electron_1974}, neutrons~\cite{zeilinger_single-_1988}, and molecules~\cite{arndt_waveparticle_1999}.
	In all these cases, the effect of diffractive focusing manifests itself~\cite{Diffractive-Focusing-General} provided the initial wave function is ($i$) a real-valued one and ($ii$) has a non-Gaussian shape. Moreover, this kind of focusing can be very useful for the collimation of waves, such as water waves and $x$-rays, for which no ordinary lenses exist. 
	
	In contrast to the studies mentioned above, in the present article we explore this phenomenon for an atomic BEC~\cite{cornell_nobel_2002,ketterle_nobel_2002} in a regime where the atom-atom interaction plays a key role. Our paper has a twofold objective: ($i$) We generalize the diffractive focusing effect to the case of {\it interacting waves}, and ($ii$) we demonstrate that, in this regime, a rather straightforward implementation is possible. 
	
	For this purpose, we consider an atomic BEC with a rectangular initial wave function emulating the case of previous studies that analyzed matter wave diffraction out of a rectangular slit \cite{case_diffractive_2012, goncalves_single-slit_2017, weisman_diffractive_2017}. 
	In the laboratory, such shapes can be realized with BECs confined to so-called box potentials leading to uniform ground-state-density distributions due to the repulsive atom-atom interactions. The required potentials can for instance be generated by blue-detuned light sheets, in some cases combined with higher-order Laguerre-Gaussian (LG) laser beams~\cite{meyrath_bose-einstein_2005, jaouadi_bose-einstein_2010, gaunt_bose-einstein_2013}. 
	
	When the trap is switched-off, the freely evolving rectangular matter wave undergoes diffractive focusing in complete analogy to a matter wave being diffracted by a rectangular aperture in the near-field or paraxial regime. The particularity of the realization we are discussing in this study is that there is no need for a dedicated aperture, but the box potential itself acts as the aperture and forms the required non-Gaussian initial state. Hence, the size of the aperture is given by the characteristic length of the box potential or by the size of the BEC itself.
	
	This straightforward implementation of diffractive focusing occurs in all spatial dimensions where the initial wave function is close to having a rectangular shape. For the sake of simplicity, we study this effect in a quasi-1D BEC configuration that can easily be generalized to 3D. Our analytical and numerical methods, based on the Gross-Pitaevskii equation (GPE) \cite{gross_structure_1961, pitaevskii_vortex_1961} as well as the dynamics of Wigner functions in phase space~\cite{schleich_quantum_2001}, help to identify the optimal conditions for observing this type of focusing and for defining several benchmarks such as the focusing factor or the focus time.
	
	We emphasize that our results apply not only to the physics of Bose-Einstein condensates \cite{pitaevskii_bose-einstein_2003}, but also to other physical systems, whose dynamics is governed by the so-called cubic Schr\"odinger equation, for example, to nonlinear fiber optics \cite{agrawal_2012} and deep water surface water waves of moderate steepness \cite{mei_1983}. In order to study the effect of the interaction, that is the magnitude of the nonlinear term in the cubic Schr\"odinger equation, we propose here to observe diffractive focusing of interacting waves with a uniform BEC. Indeed, for this system all required experimental techniques already exist.

	Our article is organized as follows. In Section 2, we derive the effective 1D GPE starting from the 3D GPE and introduce an optical potential by using a laser in a higher-order LG mode. We then focus in Section 3 on finding the optimal parameters for this potential to obtain a nearly rectangular ground state. Moreover, we study in detail the effect of the atom-atom interaction on the main features of diffractive focusing. Furthermore, we compare the results of our quasi-1D model to a full 3D treatment to prove their validity. 
	In Section 4, we explain the effect of diffractive focusing in the interacting case by studying the dynamics of the Wigner function in phase space. We conclude in Section 5 by summarizing our results and discussing further interesting avenues.  
	
	Detailed calculations are presented in three Appendices. In \ref{Appendix_derivation_of_1D_GPE} we derive the effective 1D GPE for the longitudinal wave function and the effective 1D interaction strength in the two limits of almost non-interacting and weakly interacting BECs. We then evaluate in \ref{sec:App_B_chemPot_energy} the chemical potential and the energy of a BEC in the relevant external potentials. Furthermore, \ref{fidelity_rect} presents the Thomas-Fermi wave function for the optical trapping potential.

	\section{Theoretical foundations}
	
	In order to demonstrate the effect of diffractive focusing with a BEC we consider a quasi-1D setup that contains all relevant aspects and allows for an elementary presentation of the core features. To this end, we effectively freeze the dynamics in two dimensions, and analyze the focusing of an appropriately shaped wave function in the third dimension. 
	
	In this section we first introduce the effective 1D GPE that governs the dynamics of the BEC. We then present a special form of the box-shaped trapping potential based on a higher order LG mode.

	\subsection{From a 3D to a quasi-1D Bose-Einstein condensate}
	
	To arrive at a quasi-1D BEC consisting of $N$ atoms of mass $m$, we start from the 3D GPE \cite{pitaevskii_bose-einstein_2003}
	\begin{equation}
		\imag \hbar \pdiff[]{}{t} \psi\qty{\mathbf{r}, t} = \sqty{-\frac{\hbar^{2}}{2m} \pdiff[2]{}{\mathbf{r}} + V\qty{\mathbf{r}, t} + g N \abs{\psi\qty{\mathbf{r}, t}}^{2}}\psi\qty{\mathbf{r}, t}
		\label{eq_TB_GPE_full3D}
	\end{equation}
	for the macroscopic wave function $\psi=\psi\qty{\mathbf{r}, t}$ which is normalized according to the condition 
	\begin{equation}
		\int \dd^3 {r} \abs{\psi\qty{\mathbf{r}, t}}^{2} = 1,
		\label{eq_normalisation_condition}
	\end{equation}
	where $ \mathbf{r} \equiv \qty{x,y,z} $ is the position vector with the Cartesian coordinates $x$, $y$ and $z$.

	Here we assume that the atoms are interacting via a contact potential whose strength is determined by the $s$-wave scattering length $a_s$, resulting in the interaction constant
	\begin{equation}
		g \equiv \frac{4 \pi \hbar^{2} a_s}{m}.
		\label{g3D}
	\end{equation}
	In addition, the external potential
	\begin{equation}
		V\qty{\mathbf{r}, t} \equiv V_{\perp}\qty{x,y} + V_{\mrm{Box}}\qty{z,t}
	\end{equation}
	consists of the harmonic trap 
	\begin{equation}
		V_{\perp}\qty{x,y} \equiv \frac{1}{2} m \qty{\omega_{x}^{2} x^{2} + \omega_{y}^{2} y^{2}}
		\label{eq_transverspotential}
	\end{equation}
	in the transverse directions determined by the trap frequencies $\omega_{x}$ and $\omega_{y}$, as well as the box potential $V_{\mrm{Box}}=V_{\mrm{Box}}(z,t)$ along the $z$-axis enforcing a rectangular ground state. 
	
	Throughout this paper we consider the case where the longitudinal characteristic length $L_z$ of the external potential is much larger than the transverse one $L_{\perp}\equiv \sqrt{\hbar/m\omega_{\perp}}$ with $\omega_{\perp} \equiv \sqrt{\omega_{x}\omega_{y}}$. 
	In \ref{Appendix_derivation_of_1D_GPE} and \ref{sec:App_B_chemPot_energy} we show that in this limit there is no dynamics in the transverse direction, that is in the $x$-$y$ plane, as long as $N a_s \ll L_z$. 
	
	Based on these assumption we derive in \ref{Appendix_derivation_of_1D_GPE} the effective 1D GPE \cite{couto_effective_2018}
	\begin{equation}
		\label{1D GP effective}
		\imag \hbar \pdiff[]{}{t} \varphi\qty{z,t} = \left(-\frac{\hbar^{2}}{2m} \pdiff[2]{}{z} + V_{\mrm{Box}}(z) + \tilde{g}|\varphi|^2\right)\varphi\qty{z,t}
	\end{equation}
	for the wave function $\varphi = \varphi(z,t)$ along the $z$-direction. 
	Here the effective interaction strength
	\begin{equation}
		\tilde{g} \equiv g N c_{\perp}
		\label{dimension_g}
	\end{equation}
	is determined by the interaction constant $g$, Eq.~\eqref{g3D}, the number of particles $N$, and the coupling parameter $c_{\perp}$, which in general is not a constant and originates from the non-linear coupling between the transverse ($x$-$y$ plane) and the longitudinal ($z$-axis) dynamics. 
	
	In \ref{Appendix_derivation_of_1D_GPE} we derive also analytical expressions for $c_{\perp}$ in two limiting cases, namely for ($i$) almost non-interacting and ($ii$) weakly interacting atoms. 
	If $0\leq N a_{s}\ll L_{\perp} \ll L_z$, the atom-atom interaction is so small that there is no coupling between the dynamics in the transverse and the longitudinal degrees of freedom, resulting in the expression
	\begin{equation}
		\label{c_perp-weak}
		c_{\perp} = \frac{1}{2\pi L_{\perp}^2}.
	\end{equation}

	In the opposite limit, if $L_{\perp} \ll N a_{s} \ll L_z$, we can apply the Thomas-Fermi approximation \cite{pitaevskii_bose-einstein_2003} for the ground state of the 3D GPE, Eq.~\eqref{eq_TB_GPE_full3D}, and arrive at the formula
	\begin{equation}
		\label{c_perp-strong}
		c_{\perp} = \frac{1}{2\pi L_{\perp}^2} \left(\frac{8}{9}\frac{L_z}{Na_s}\right)^{\frac{1}{2}}
	\end{equation}
	for the parameter $c_{\perp}$.
	
	We conclude by noting that in section \ref{3s simulation} we perform full 3D numerical simulations based on Eq.~\eqref{eq_TB_GPE_full3D}, to test the validity of the effective 1D description. We find that for the parameters considered in this article the 1D GPE, Eq.~\eqref{1D GP effective}, with the coupling parameter $c_\perp$ given by Eq.~\eqref{c_perp-strong} describes correctly the dynamics of the BEC.

	\subsection{Optical box potential}
\begin{figure}[t]
		\centering
		\includegraphics[width=10cm]{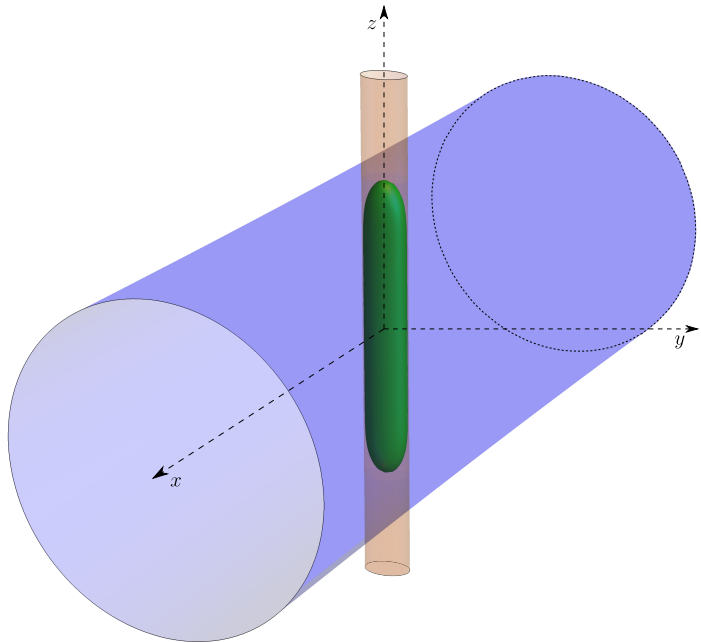}
		\caption{Trap arrangement for diffractive focusing of a uniform BEC. 
			The harmonic trap $V_{\perp} \equiv V_{\perp}(x,y)$, Eq.~\eqref{eq_transverspotential}, (orange) and the LG potential $V_l=V_l(\sqrt{y^2+z^2})$, Eq.~\eqref{eq_full_LG_potential}, (blue) cause a confinement for the atoms in the transverse ($x$-$y$ plane) and the longitudinal ($z$-axis) directions, respectively, yielding a cigar-shaped ground-state-density distribution (green) of the BEC.}
		\label{figure_3D_potential}
	\end{figure}

	We realize the box potential $V_{\mrm{Box}}$ by a LG laser beam, more precisely by the radially symmetric ${\rm LG}_0^l$ mode with the intensity profile \cite{jaouadi_bose-einstein_2010}
	\begin{equation}
		\label{eq_LG_intensity}
		I_l(\rho)=\frac{2}{\pi l!} \frac{P}{w_0^2} \left( \frac{2\rho^2}{w_0^2} \right)^l \exp \left( -\frac{2\rho^2}{w_0^2} \right),
	\end{equation}
	where $l=0,1,2,...$ is the order of the mode, $w_0$ and $P$ are the waist and the power of the beam, respectively. Here $\rho \equiv \sqrt{y^2 + z^2}$ measures the distance from the beam axis, as depicted in Fig. \ref{figure_3D_potential}.
	
	By choosing the waist $w_0$ of the LG beam to be much larger than the characteristic lengths of the harmonic trap along the transverse directions, that is when $w_0 \sqrt{l}\gg L_{\perp}$, as shown in Fig. \ref{figure_3D_potential}, we can neglect the cylindrical symmetry of the beam profile and use the effective 1D intensity profile $I_l(\rho = z)$ along the $z$-axis instead. 
	
	In addition, if the laser frequency is far blue-detuned from the atomic resonance, the associated optical dipole potential reads \cite{jaouadi_bose-einstein_2010, GRIMM200095}
	\begin{equation}
		\label{eq_full_LG_potential_intensityrelation}
		V_{l}(z)=\frac{\hbar\Gamma^2}{8\Delta}\frac{I_l(z)}{I_s}.
	\end{equation}
	Here $\Gamma$, $\Delta$ and $I_s$ denote the decay rate, the detuning, and the saturation intensity, respectively.
	
	With the help of the explicit expression, Eq.~\eqref{eq_LG_intensity}, of the intensity, we find with $I_l(\rho = z)$ the formula
	\begin{equation}
		\label{eq_full_LG_potential}
		V_{l}(z)=\frac{2^l}{4\pi l!}\frac{\hbar\Gamma^2}{\Delta}\frac{P}{I_s w_0^2}
		\left(\frac{z}{w_0}\right)^{2l}\exp\left(-\frac{2z^2}{w_0^2}\right)
	\end{equation}
	for the trapping potential caused by the LG mode.
	
	If the chemical potential of the ground state is much lower than the maximum $V_l(z_l)$ of the trapping potential located at $z_l\equiv w_0\sqrt{l/2}$, we can approximate Eq.~\eqref{eq_full_LG_potential} around the potential minimum at $z=0$ and obtain the power-law
	\begin{equation}
		\label{eq_pep_of_initial_state__approximatepotential}
		V_l(z)\cong \frac{2^l}{4\pi l!}\frac{\hbar \Gamma^2}{\Delta}\frac{P}{I_s w_{0}^2}\left(\frac{z}{w_{0}}\right)^{2l}
	\end{equation}
	for the trapping potential $V_l$. 
	
	In the following we choose our parameters such that this power-law approximation is valid and Eq.~\eqref{eq_pep_of_initial_state__approximatepotential} can be used to describe the box potential. Moreover, we note that similar box-like trapping potentials can also be realized by combining appropriate Hermite-Gauss beams with Gauss beam endcaps \cite{meyrath_bose-einstein_2005}, or by employing blue-detuned painted potentials \cite{gaunt_bose-einstein_2013}.

	\section{Diffractive focusing}
	\label{sec_results}
	In this section we first identify the parameters of the LG potential $V_l$, Eq.~\eqref{eq_pep_of_initial_state__approximatepotential}, which allow us to create a quasi-1D BEC ground state wave function with the desired rectangular shape. By taking this state as the initial one, we then study in detail the effect of the atom-atom interaction on the diffractive focusing. Finally, we compare the results of our quasi-1D model to a full 3D treatment.
	
	\subsection{Preparation of the initial state of rectangular shape}
	\label{subsection-initial state}
	
	We start with the discussion of the ground state of a quasi-1D BEC in the LG potential given by Eq.~\eqref{eq_pep_of_initial_state__approximatepotential}. To be specific, we throughout our article consider $^{87}\mathrm{Rb}$ atoms \cite{van_kempen_interisotope_2002, daniel_steck_rubidium87} and emphasize that the values of all relevant parameters, listed in Table \ref{tab:values}, are accessible in a state-of-the art experiment. 
	
	\begin{table}[t]
		\centering
		\caption{Parameters and their values used in our numerical simulations based on $^{87}\mathrm{Rb}$ atoms. Here $a_0$ denotes to the Bohr radius.}
		\label{tab:values}
		\begin{tabular}{lcr}
			\\
			\hline 
			\hline
			number of atoms & $N$ &  $ 10^4 $\\ 
			scattering length & $a_s$ &  $ 100\, a_0$\\ 
			trap frequencies & $\omega_{\perp}=\omega_{x}=\omega_{y}$ & $ 2 \pi \cdot 2.5\cdot 10^3\, \mathrm{Hz} $ \\ 
			transverse length & $L_\perp$ & $2.16 \cdot 10^{-7}\,\mathrm{m}$ \\
			longitudinal length, beam waist & $L_{z}\cong w_0$ &  $15\cdot 10^{-6}\,\mathrm{m}$\\ 
			effective frequency & $\omega_z = \frac{\hbar}{m L_{z}^2}$ &  $2\pi \cdot 0.52\,\mathrm{Hz}$\\ 
			detuning & $\Delta$ & $ 2\pi \cdot 1.0\cdot 10^{13}\, \mathrm{Hz} $ \\ 
			decay rate & $\Gamma$ & $ 2 \pi \cdot 6.1 \cdot 10^6\, \mathrm{Hz} $ \\ 
			saturation intensity & $I_s$ & $ 16 \,\mathrm{W}\cdot{\rm m^{-2}} $ \\ 
			laser power & $P$ &  $0.1\,\mathrm{W}$ \\ 
			\hline
			\hline
		\end{tabular}
	\end{table}

	According to Table \ref{tab:values} we find for the ratio
	\begin{equation}
		\frac{N a_{s}}{L_{\perp}} \cong 245, 
		\label{eq_condition_TF_fulfiled}
	\end{equation}
	which implies that the atoms are indeed weakly interacting. 
	
	Since in this case the parameter $c_{\perp}$ is given by Eq.~\eqref{c_perp-strong}, we obtain from Table \ref{tab:values} the value
	\begin{equation}
		c_{\perp} \cong 1.7 \cdot 10^{12}\;\mathrm{m}^{-2}.
		\label{eq_correction_factor}
	\end{equation}

\begin{figure}[t]
		\centering
		\includegraphics{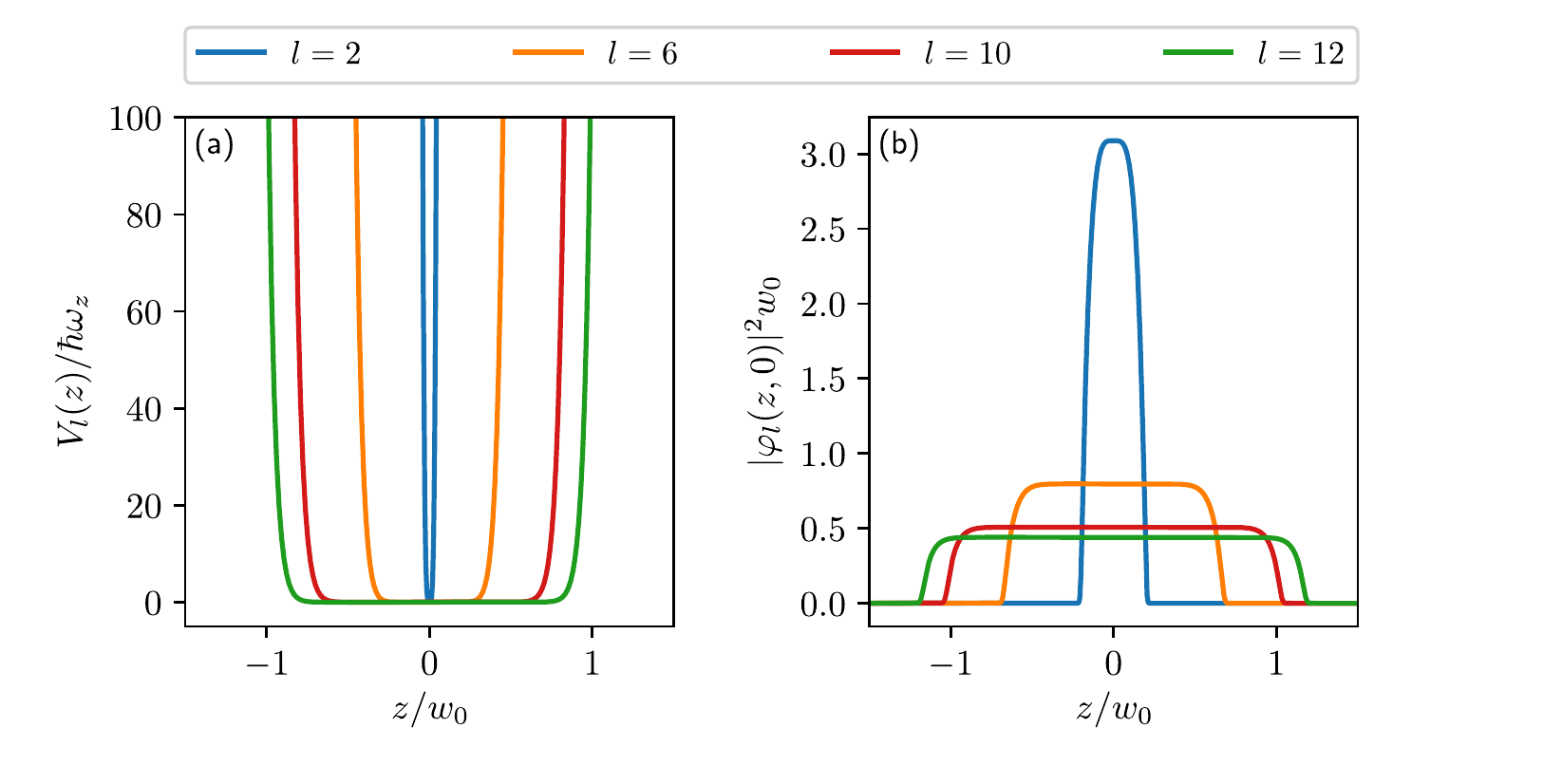}
		\caption{Creation of a rectangular wave function of an interacting BEC with the help of an optical box potential. The LG potential $V_l=V_l(z)$ (a) of mode order $l$ is given by Eq.~\eqref{eq_pep_of_initial_state__approximatepotential} and the normalized probability density $P_l\equiv |\varphi_l(z,0)|^2$ (b) of the corresponding ground state is calculated numerically for the values of $l=2,~6,~10$ and $12$. For values $l \geq 10$ a sufficient rectangularity of the ground state density distribution is reached, as show in Fig. \ref{appendig_C_overlap} }
		\label{figure_groundstates_of_lgpotentials}
	\end{figure}

	With the help of the imaginary-time propagation method \cite{auer_fourth-order_2001} we have solved Eq.~\eqref{1D GP effective} numerically, and have obtained the wave function $\varphi_l = \varphi_l(z,t)$ for the ground state of a BEC in the LG box potential $V_l$ given by Eq.~\eqref{eq_pep_of_initial_state__approximatepotential} for different values of $l$, namely $l=2,6,10,12$, as shown in Fig.~\ref{figure_groundstates_of_lgpotentials}.
	
	For increasing values of $l$ the potential $V_l$ becomes more rectangular, that is flatter at $z=0$ and steeper at the edges as displayed in Fig.~\ref{figure_groundstates_of_lgpotentials} (a). Since for the ground state we consider the natural scattering length of $^{87}\mathrm{Rb}$ we fulfill the condition of the Thomas-Fermi approximation \cite{pitaevskii_bose-einstein_2003} and the ground-state wave function has the form of the inverse potential, as shown in \ref{fidelity_rect}. Thus, when the potential gets more rectangular, the ground state is more rectangular, too, as apparent from Fig. \ref{figure_groundstates_of_lgpotentials} (b), because the interaction between the atoms enforces a more homogeneous density distribution within the box potential. 
	
	Indeed, the fidelity 
	\begin{equation}
	    \mathcal{F} = \int_{-\infty}^\infty \mathrm{d} z \,\varphi_l (z) \varphi_l^{(R)}(z)
	    \label{calculationOfOverlap}
	\end{equation}
	between the ground state wave function $\varphi_l$ and the normalized wave function 
	\begin{equation}
	    \varphi_l^{(R)}(z) = \sqrt{h_l}\, \Theta\left(\frac{1}{2h_l} - |z|\right)
	    \label{rectfunc_overla}
	\end{equation}
	of a rectangular shape with the same amplitude $h_l\equiv\max(\varphi_l)$, serves as our measure of the rectangularity of the ground state.	
	
\begin{figure}
		\centering
		\includegraphics[width=0.75\columnwidth]{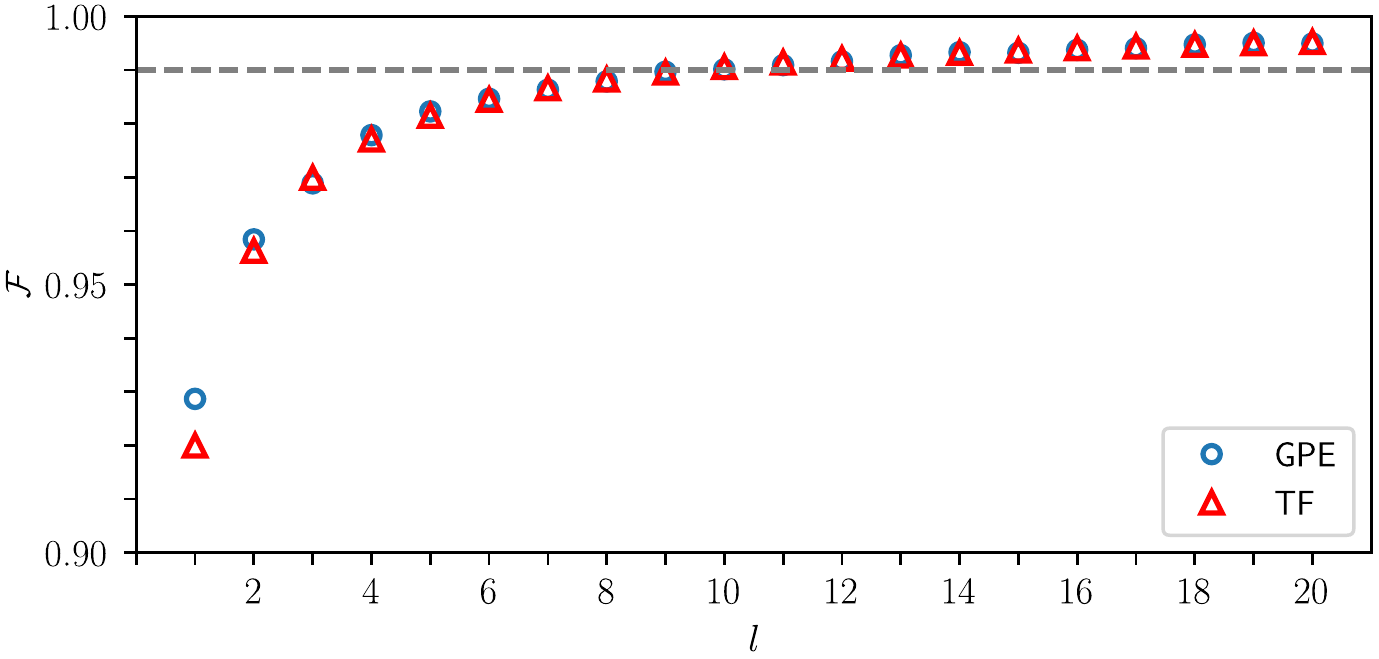}
		\caption{Determination of the optimal order $l$ of the LG mode with the help of the fidelity $\mathcal{F}$, Eq.~\eqref{calculationOfOverlap}. Here we obtain the overlap between the rectangular wave function $\varphi_l^{(R)}(z)$, Eq.~\eqref{rectfunc_overla}, and $\varphi_l$ given by either the Thomas-Fermi wave function $\varphi_{TF}$, Eq.~ \eqref{phi-TF}, (red triangles) or the ground state wave function (blue dots) obtained by solving numerically the GPE, Eq. \eqref{1D GP effective}, in its dependence on the values of $l$ of the potential $V_l$, Eq. \eqref{eq_pep_of_initial_state__approximatepotential}. The gray dashed line marks the level of $99\%$ for $\mathcal{F}$.}
		\label{appendig_C_overlap}
	\end{figure}

	As displayed in Fig.~\ref{appendig_C_overlap} for the fidelity ${\mathcal F}$ as a function of $l$, for $l\geq  10$ the fidelity reaches 99\% and we therefore consider the case $l=10$ in the remainder of our article.

	\subsection{Influence of atom-atom interaction }
	
	Now we are in the position to analyze the effect of the atom-atom interaction on diffractive focusing. We start with the ground state of the BEC in the trapping potential $V_{\perp}+V_{10}$ and then switch off only the LG potential $V_{10}$ while simultaneously changing the scattering length $a_s$ from its initial value, $a_s=100\,a_0$, to its final one $a_{s}^{\rm (f)}$ via a Feshbach resonance \cite{chin_feshbach_2010}, where $a_0$ corresponds to the Bohr radius.
	
\begin{figure}[t]
		\centering
		\includegraphics{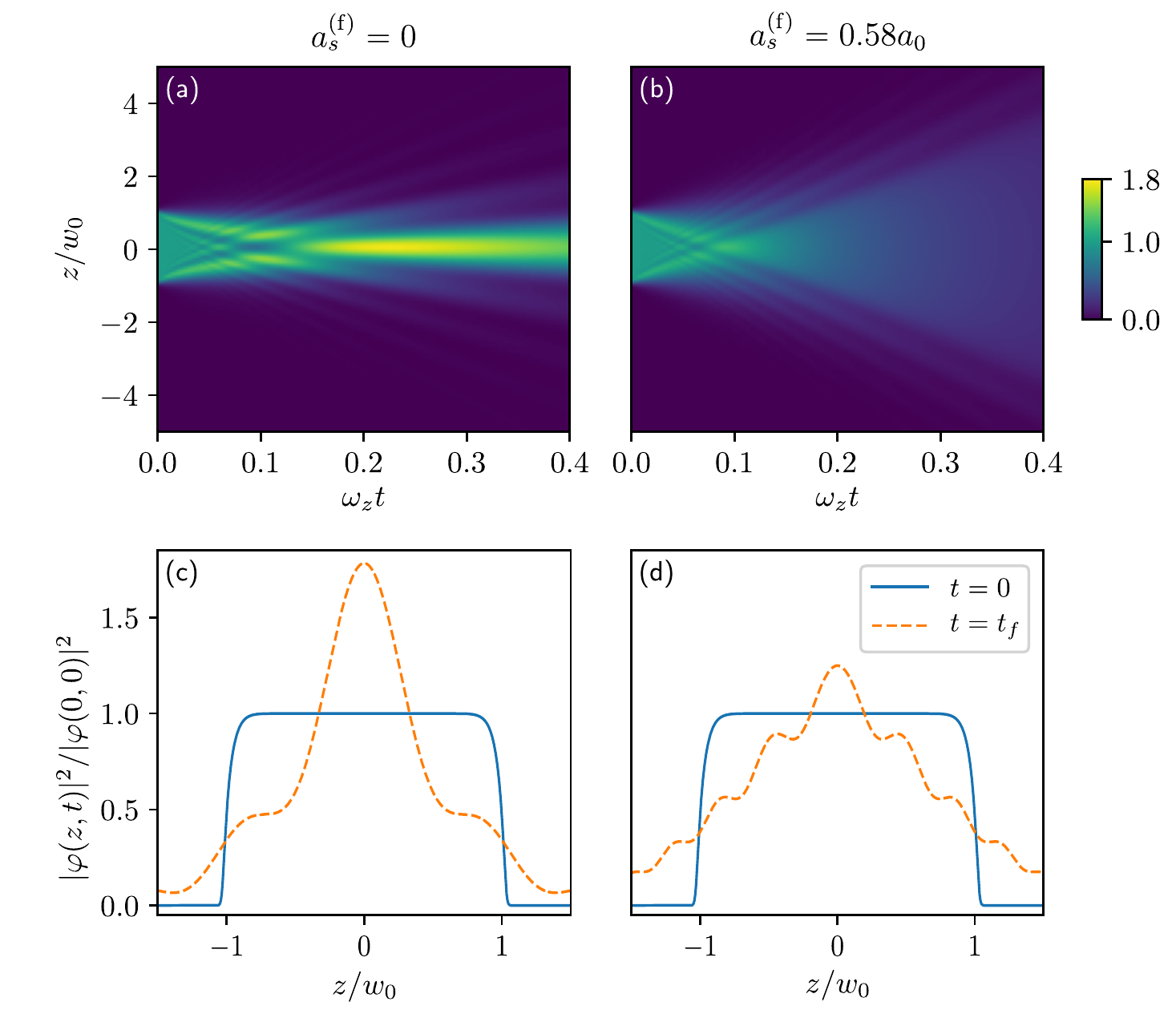}
		\caption{Influence of the atom-atom interaction on the phenomenon of diffractive focusing of a uniform BEC apparent in the time evolution of the normalized 1D density distribution $|\varphi(z,t)|^2/|\varphi(0,0)|^2$. The initial state $\varphi(z,0)$ is the ground state of the trapping potential $V=V_{\perp}+V_{10}$ and the dynamics (a, b) takes place after switching off only the longitudinal potential $V_{10}$ {\it and} changing the scattering length $a_s$ from its initial value $a_s=100\,a_0$ to its final one $a_s^{\rm (f)}=0$ (a), or $a_s^{\rm (f)}=0.58\,a_0$ (b). The bottom row (c, d) presents the corresponding density distributions at $t=0$ (blue line) and at the focal time $t=t_f$ (orange dashed line). For both choices of the scattering length the focusing appears at $z=0$ for the focal times $t_f=0.22/\omega_z$ and $t_f=0.09/\omega_z$ with the focusing factors $f=1.78$ and $f=1.25$.}
		\label{fig_free_time_evo_gdyn_100_densityplot}
	\end{figure}

	The resulting time evolution in the $z$-direction, calculated from Eq.~\eqref{1D GP effective} with a split-step algorithm \cite{javanainen_symbolic_2006}, is shown in Fig. \ref{fig_free_time_evo_gdyn_100_densityplot} for two different values of the final scattering length, namely for $a_s^{\rm (f)} = 0$ (a, c) and $a_s^{\rm (f)} = 0.58\,a_{0}$ (b, d). The maximum of the distribution appears at $z=0$ in both cases.
	
	Since the initial wave function $\varphi(z,0)$ has a nearly rectangular shape, we characterize the focusing effect by the focusing factor 
	\begin{equation}
		f \equiv \frac{\max\limits_{t} \abs{\varphi(0,t)}^{2}}{\max\limits_{z} \abs{\varphi(z,0)}^{2}},
		\label{f-definition}
	\end{equation}
	which describes the increase of the amplitude of the probability density during the dynamics in comparison with its initial value.
	
	From Fig. \ref{fig_free_time_evo_gdyn_100_densityplot}, we note the factors $f=1.78$ and $f=1.25$ for $a_s^{\rm (f)}=0$ and $a_s^{\rm (f)}=0.58\, a_{0}$, respectively. Moreover, the repulsive atom-atom interaction ($a_s^{\rm (f)}>0$) results in ($i$) a decrease of the focal time $t_f$, and ($ii$) a faster spreading of the wave function directly after the focus.

	\begin{figure}[t]
		\centering
		\includegraphics{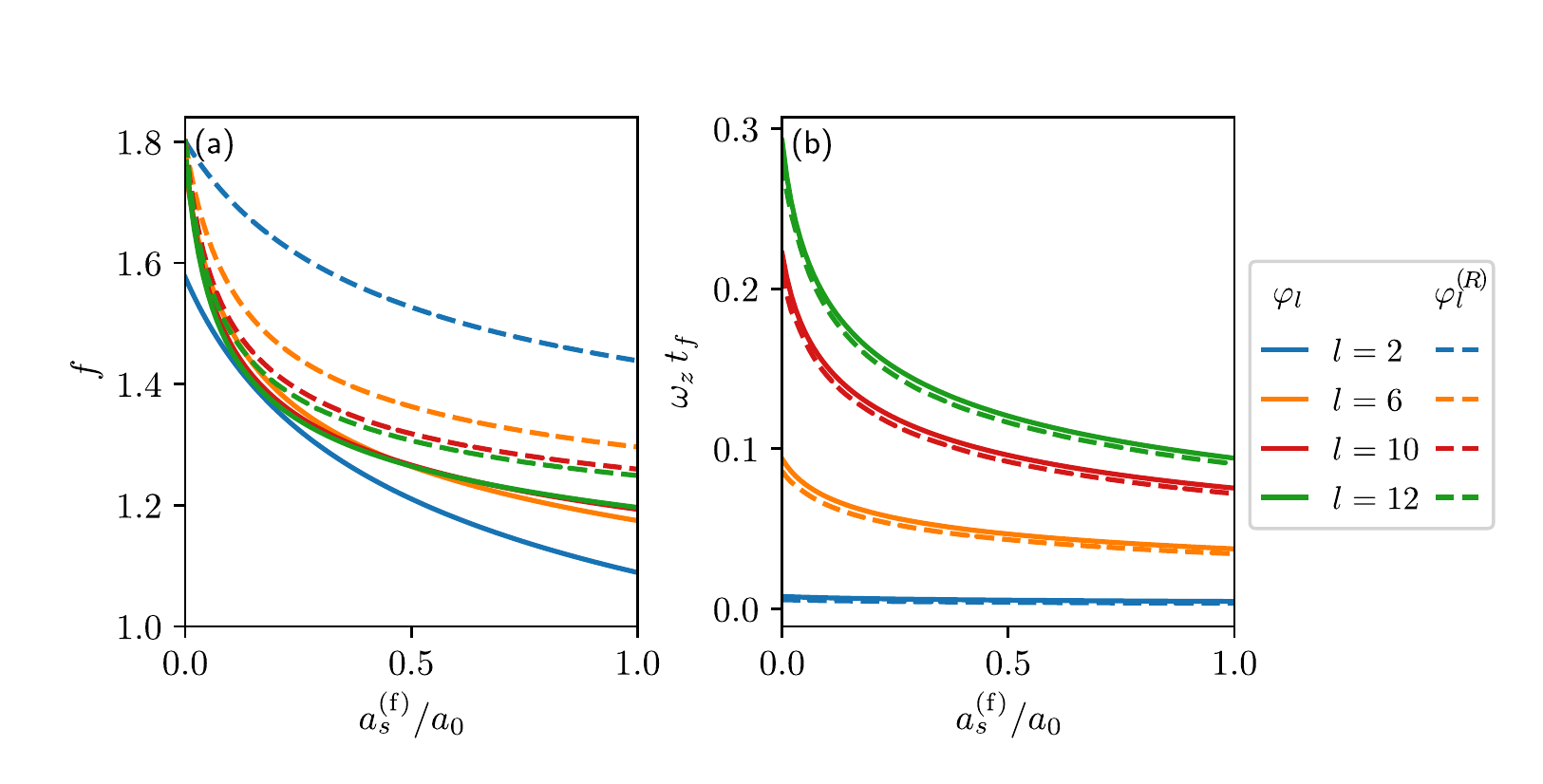}
		\caption{Dependence of the focusing factor $f$ (a) and the focal time $t_f$ (b) of a uniform BEC on the final scattering length $a_s^{\rm (f)}$ for different orders $l$ of the LG mode, with the initial state characterized by the ground state wave function $\varphi_l$ of the potential $V_{\perp}+V_l$ (solid lines) or the rectangular wave function $\varphi_l^{(R)}$ (dashed lines) with the same height, as defined by Eq.~\eqref{rectfunc_overla}. For a growing final scattering length $a_s^{\rm (f)}$ the focusing factor $f$ and the focal time $t_f$ both decrease rapidly.}
		\label{fig_focusfactor_of_dynamical_interaction_diffl}
	\end{figure}

	Figure \ref{fig_focusfactor_of_dynamical_interaction_diffl} displays the dependence $f = f\left(a_s^{\rm (f)}\right)$ and $t_f = t_f\left(a_s^{\rm (f)}\right)$ for four different values of $l$, with the initial state being the ground state of the complete potential $V=V_{\perp}(x,y)+V_l(z)$ (solid lines), or the corresponding rectangular state (dashed lines), defined by Eq.~\eqref{rectfunc_overla}, respectively. For a growing interaction strength the focusing factor and the focal time decrease rapidly. Hence, diffractive focusing is only visible if the atom-atom interaction is very weak during the dynamics. 
	
	Figure \ref{fig_focusfactor_of_dynamical_interaction_diffl} (b) shows that for a given $l$ the focal time of the system depends strictly on the size of the initial profile, and the results for the ground-state wave function resemble the ones for the corresponding rectangular state. 
	
	One the other hand, Fig. \ref{fig_focusfactor_of_dynamical_interaction_diffl} (a) shows that a appropriate fidelity, as defined by Eq. \eqref{calculationOfOverlap}, is necessary to reduce the deviations in the predictions of the focus factors corresponding to $\varphi_l$ and $\varphi_l^{(R)}$. This difference can be explained by the fact that the edges cover a larger area of the wave function, compared to the flat surface, for small values of $l$ than for large values of $l$. For our studies we have used the ground-states $\varphi_{l}(z,0)$ of $l=10$ (red curve), which provides a fidelity of more than $99\%$, as shown in Fig. \ref{appendig_C_overlap}.

	\begin{figure}[t]
		\centering
		\includegraphics{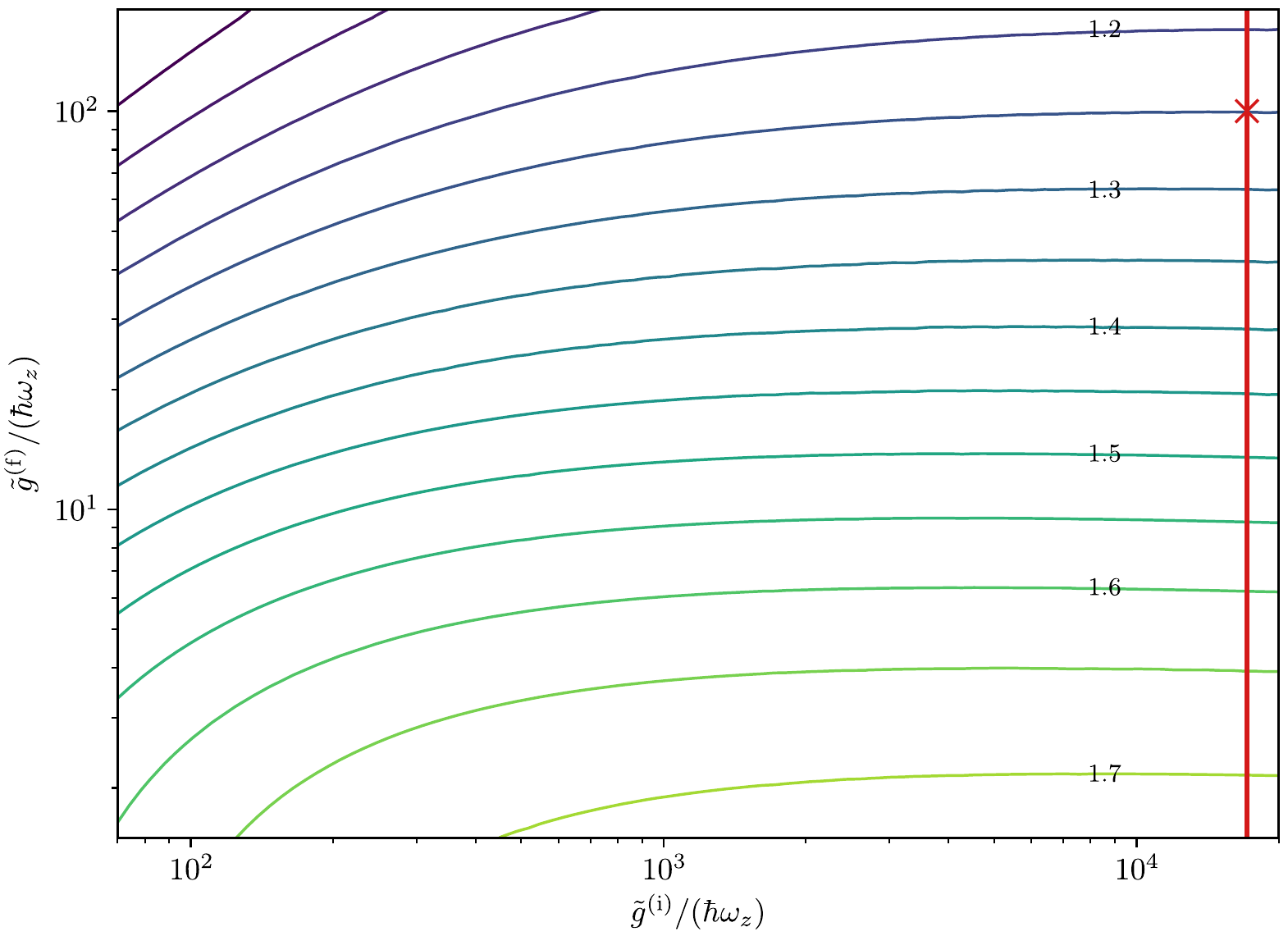}
		\caption{Contour plot of the focusing factor $f$ as a function of the initial and final effective interaction strength $\tilde{g}^{(i)}$ and $\tilde{g}^{(\rm f)}$. The focusing factor $f$ depends only weakly on the {\it initial} interaction $\tilde{g}^{(\rm i)}$. However, a vertical cut (red line) through the contour plot at a fixed value of $\tilde{g}^{(i)} = 17129.71 \,\hbar \omega_z$ corresponding to $a_s=100\,a_0$ (red line) reveals that $f$ depends strongly on the {\it final} interaction length $\tilde{g}^{(\rm f)}$, in complete agreement with the dependence shown in Fig.~\ref{fig_focusfactor_of_dynamical_interaction_diffl} (a). The reference case from Fig.~\ref{fig_free_time_evo_gdyn_100_densityplot} (b,d) is marked by a red cross. The corresponding scattering length giving rise to the interaction strength $\tilde{g}$ can be calculated by inverting Eq.~\eqref{dimension_g} and choosing an appropriate value of $c_\perp$.}
		\label{parameterscan_diffinitialpotential_dynamicinteraction}
	\end{figure}
	
	Furthermore, Fig.~\ref{parameterscan_diffinitialpotential_dynamicinteraction} presents a contour plot of the focusing factor $f$ for different values of the initial and final effective interaction strength, Eq.~\eqref{dimension_g}, $\tilde{g}^{(i)}$ and $\tilde{g}^{(\rm f)}$ with $l=10$. For small values of $\tilde{g}^{(i)} \rightarrow 0$ ($a_s\rightarrow0$) the focusing effect reduces drastically, as the ground state wave function approaches the one of a particle in the box potential, that is the non-Gaussian shape vanishes. 
	
	On the other hand, we recognise that values of $\tilde{g}^{(i)} = 3000\, \hbar \omega_z$ already result in similar behaviour as our reference case of $\tilde{g}^{(i)} = 17129.71 \,\hbar \omega_z$, which corresponds to $a_s = 100\,a_0$.
	Here the interaction $\tilde{g}$ is used since we undergo a transition of the two limiting cases (i) almost non-interacting (small $\tilde{g}$) and (ii) weakly interacting atoms (large $\tilde{g}$), and in general the relation between $\tilde{g}$ and $a_s$ is unknown.
	Moreover, for any values $\tilde{g}^{(i)}>0$, the focusing factor $f$ decreases for growing values of $\tilde{g}^{\rm (f)}$, as displayed in Figs. \ref{fig_focusfactor_of_dynamical_interaction_diffl} and \ref{parameterscan_diffinitialpotential_dynamicinteraction}.
	
	To summarize, in order to observe the phenomenon of diffractive focusing for a quasi-1D BEC, we require ($i$) a large initial interaction $\tilde{g}^{(i)}$ for preparing a nearly rectangular state, and ($ii$) a small final interaction $\tilde{g}^{\rm (f)}$ to have a measurable effect during the dynamics. Hence, experimentally the use of Feshbach resonances \cite{chin_feshbach_2010} is mandatory to tune the atom-atom interaction in the desired way. As mentioned at the beginning of this discussion we use $^{87}\mathrm{Rb}$, but this effect takes place for any BEC with low temperature, for example, $^{39}\mathrm{K}$ with a wide Feshbach resonance \cite{Roati_2007}.

	\subsection{Justification of the quasi-1D approximation}
	\label{3s simulation}
	
	We conclude this discussion of the phenomenon of diffractive focusing in a BEC by briefly examining to what degree the effective 1D GPE given by Eq.~\eqref{1D GP effective} describes the free propagation of the quasi-1D BEC. 
	For this purpose, we again start from the ground state of the BEC in the trapping potential $V = V_{\perp}(x,y)+V_{10}(z)$ and then switch off the LG potential $V_{10}$, while simultaneously changing the scattering length to its final value $a_{s}^{\rm (f)}$. 
	
	\begin{figure}[t]
		\includegraphics{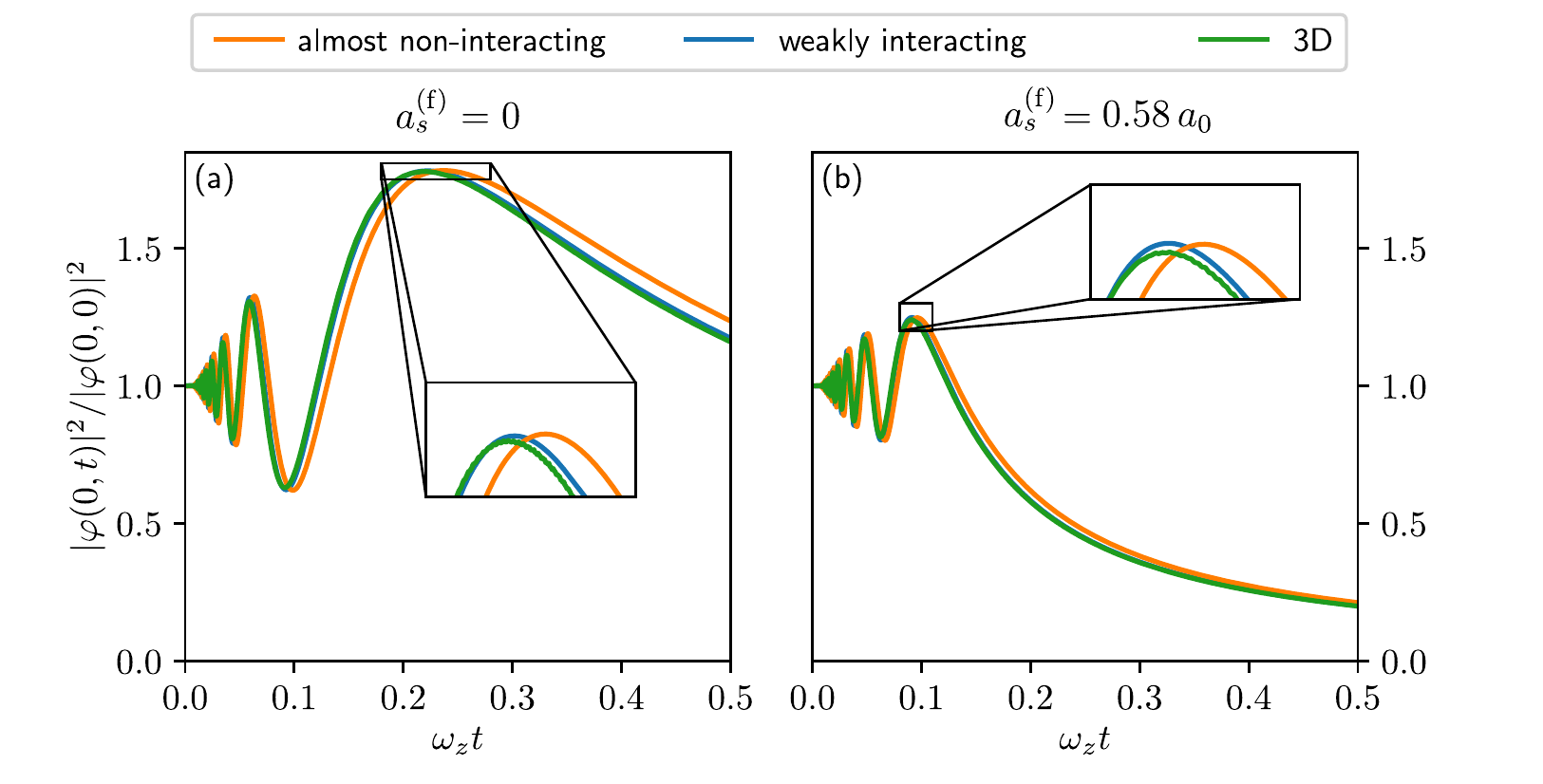}
		\caption{Time evolution of the normalized 1D density at $z=0$, with the transverse wave function $\Phi_0$ given by Eq.~\eqref{Phi0 - small g}, (orange curve), or Eq.~\eqref{Phi0-TF-result}, (blue curve). For comparison, the green curve displays the time evolution of the normalized integrated density $P_{\mathrm{3D}}(z,t)$ given by Eq.~\eqref{density_3D}, where $\psi = \psi(x,y,z,t)$ is based on the numerical solution of the 3D GPE defined by Eq.~\eqref{eq_TB_GPE_full3D}. For all cases, the initial state is given by the ground state of the trapping potential $V_{\perp}+V_{10}$ and the dynamics occurs after switching off only the longitudinal potential $V_{10}$ \textit{and} instantaneously changing the scattering length $a_s$ from its initial value $a_s=100\,a_0$ to its final one $a_s^{\rm (f)}=0$ (a), or $a_s^{\rm (f)}=0.58\,a_0$ (b).}
		\label{fig_comparison3D}
	\end{figure}

	First, we solve Eq.~\eqref{1D GP effective} for the wave function $\varphi = \varphi(z,t)$ numerically and obtain the time dependence of the normalized 1D density $|\varphi(0,t)|^2/\max\limits_{z}|\varphi(z,0)|^2$ at $z=0$. Here we consider the two cases of almost non-interacting and weakly interacting atoms depicted in Fig. \ref{fig_comparison3D} by the orange and blue curve, respectively. For these limits the parameter $c_{\perp}$, Eq.~\eqref{c_perp-def_Appendix}, is determined by the transverse wave function $\Phi_0 = \Phi_0(x,y)$, Eq. \eqref{eq_TB_seperation_condition_Appendix}, and given by Eqs. \eqref{c_perp-weak} and \eqref{c_perp-strong}, respectively.
	
	Then, we perform the full 3D numerical simulation of the GPE given by Eq.~\eqref{eq_TB_GPE_full3D} for the wave function $\psi=\psi({\bf r},t)$. The time evolution of the normalized integrated density $P_{\mathrm{3D}}(0,t)/\max\limits_{z}(P_{\mathrm{3D}}(z,0))$ at $z=0$, with 
	\begin{equation}
		P_{\mathrm{3D}}(z,t) \equiv \int \dd x\,\dd y |\psi({\bf r},t)|^2,
		\label{density_3D}
	\end{equation}
	is displayed by the green curve in Fig. \ref{fig_comparison3D}.  
	
	As a result, for our trap configuration, with the relevant parameters listed in Table \ref{tab:values}, the quasi-1D approximation is very reliable and in excellent agreement with the results of the full 3D simulation.
	
	A comparison between the curves corresponding to ($i$) almost non-interacting atoms (orange line) and ($ii$) weakly interacting atoms (blue line) with the full 3D curve reveals that the ground state obtained within the Thomas-Fermi approximation and leading to the interaction parameter $c_{\perp}$ given by Eq.~\eqref{c_perp-strong}, is more accurate in describing the dynamics of the system. This statement holds true even for small values of $a_s^{(\rm f)}$, as long as the ground state was created for large values of the initial scattering length $a_s$.
	
	We note that due to the change of the scattering length at $t=0$ the transverse wave function does not describe the ground state anymore and the system undergoes collective excitations. In our 3D simulations we observed such breathing oscillations in the transverse direction. However, as a consequence of the large anisotropy of the trapping potential, the time scale of the transverse dynamics is much shorter compared to the longitudinal motion and, thus, the influence of these fast oscillations on the slower longitudinal dynamics mostly averages out for the parameters considered in this article.

	\section{Diffractive focusing viewed from Wigner phase space}
	
	This section illuminates the phenomenon of diffractive focusing of a BEC from quantum phase space. For this purpose we first recall the essential ingredients of the Wigner formulation \cite{schleich_quantum_2001} of quantum mechanics and then study classical trajectories in the absence and the presence of an atom-atom interaction. This elementary approach provides us with a deeper insight into the dynamics of the Wigner function for an interacting matter wave.

	\subsection{Wigner function essentials}
	
	The Wigner function \cite{schleich_quantum_2001} corresponding to the wave function $\varphi=\varphi(z,t)$ is defined as
	\begin{equation}
		\label{definition_wignerfunction}
		W(z,p;t)\equiv\frac{1}{2\pi\hbar}\int_{-\infty}^{\infty}dy \exp\left(\frac{i}{\hbar}py\right)\varphi^*\left(z+\frac{y}{2},t\right)\varphi\left(z-\frac{y}{2},t\right),
	\end{equation}
	where $p$ is the momentum. 
	
	Integration of $W$ over $p$, or over $z$ yields the relations
	\begin{equation}
		\label{eq:wigner_position}
		\int_{-\infty}^{\infty}\dd{p} W(z,p;t)  = |\varphi(z,t)|^{2},
	\end{equation}
	or
	\begin{equation}
		\label{eq:wigner_momentum}
		\int_{-\infty}^{\infty} \dd{z} W(z,p;t) = |\tilde{\varphi}(p,t)|^{2},
	\end{equation}
	connecting the marginals of $W$ to the probability density distributions $|\varphi(z,t)|^{2}$ and $|\tilde{\varphi}(p,t)|^{2}$ in position and momentum space, respectively \cite{schleich_quantum_2001}.
	Here 
	\begin{equation}
		\tilde{\varphi}(p,t)\equiv\frac{1}{\sqrt{2\pi\hbar}} \int_{-\infty}^{\infty} \dd{z}  \exp\left(-\frac{i}{\hbar}pz\right)\varphi(z,t)
	\end{equation}
	is the momentum representation of the wave function $\varphi = \varphi(z,t)$.
	
	Although the Wigner function $W$ is normalized, that is
	\begin{equation}
		\int_{-\infty}^{\infty} dz \int_{-\infty}^{\infty} dp\; W(z,p;t)=1, 
	\end{equation}
	these properties do not imply that $W$ is always positive. Indeed, the Wigner function is a quasi-probability distribution \cite{schleich_quantum_2001} and its negative parts reflect the quantum features of the system under consideration.

	\subsection{Classical trajectories}
	
	Instead of deriving and solving the dynamical equation for the Wigner function corresponding to the 1D GPE, Eq.~\eqref{1D GP effective}, we obtain the time-dependent Wigner function directly from the definition, Eq.~\eqref{definition_wignerfunction}, of $W$ in terms of the time-dependent wave function $\varphi = \varphi(z,t)$ determined by solving Eq.~\eqref{1D GP effective} numerically. 
	
	In order to visualize the dynamics in phase space, we take a point $\{z,p\}$ in phase space and find the "classical trajectories" $\left\{Z(t),P(t)\right\}$ governed by the Hamilton equations
	\begin{eqnarray}
		\label{eq:classicaltrajectories}
		\frac{d}{dt}Z(t) & = &\frac{\partial}{\partial P}H(Z,P;t),\\
		\label{eq:classicaltrajectories2}
		\frac{d}{dt}P(t) & = & -\frac{\partial}{\partial Z}H(Z,P;t)
	\end{eqnarray}
	subjected to the initial conditions $Z(0)\equiv z$ and $P(0) \equiv p$. Here the classical Hamiltonian
	\begin{equation}
		\label{eq:classicaltrajectories:hamiltonian}
		H(Z, P; t) \equiv \frac{P^2}{2m} + \tilde{g} |\varphi(Z,t)|^{2}
	\end{equation}
	corresponds to the 1D GPE, Eq.~\eqref{1D GP effective}, without the trapping potential. 
	
	We emphasize that the use of Eqs. \eqref{eq:classicaltrajectories}, \eqref{eq:classicaltrajectories2} and \eqref{eq:classicaltrajectories:hamiltonian} implies the knowledge of the wave function $\varphi = \varphi(z,t)$ at all times obtained by numerically solving Eq.~\eqref{1D GP effective}.

	\subsection{Time evolution without atom-atom interaction}
	
\begin{figure}
		\centering
		\includegraphics[width=0.75\columnwidth]{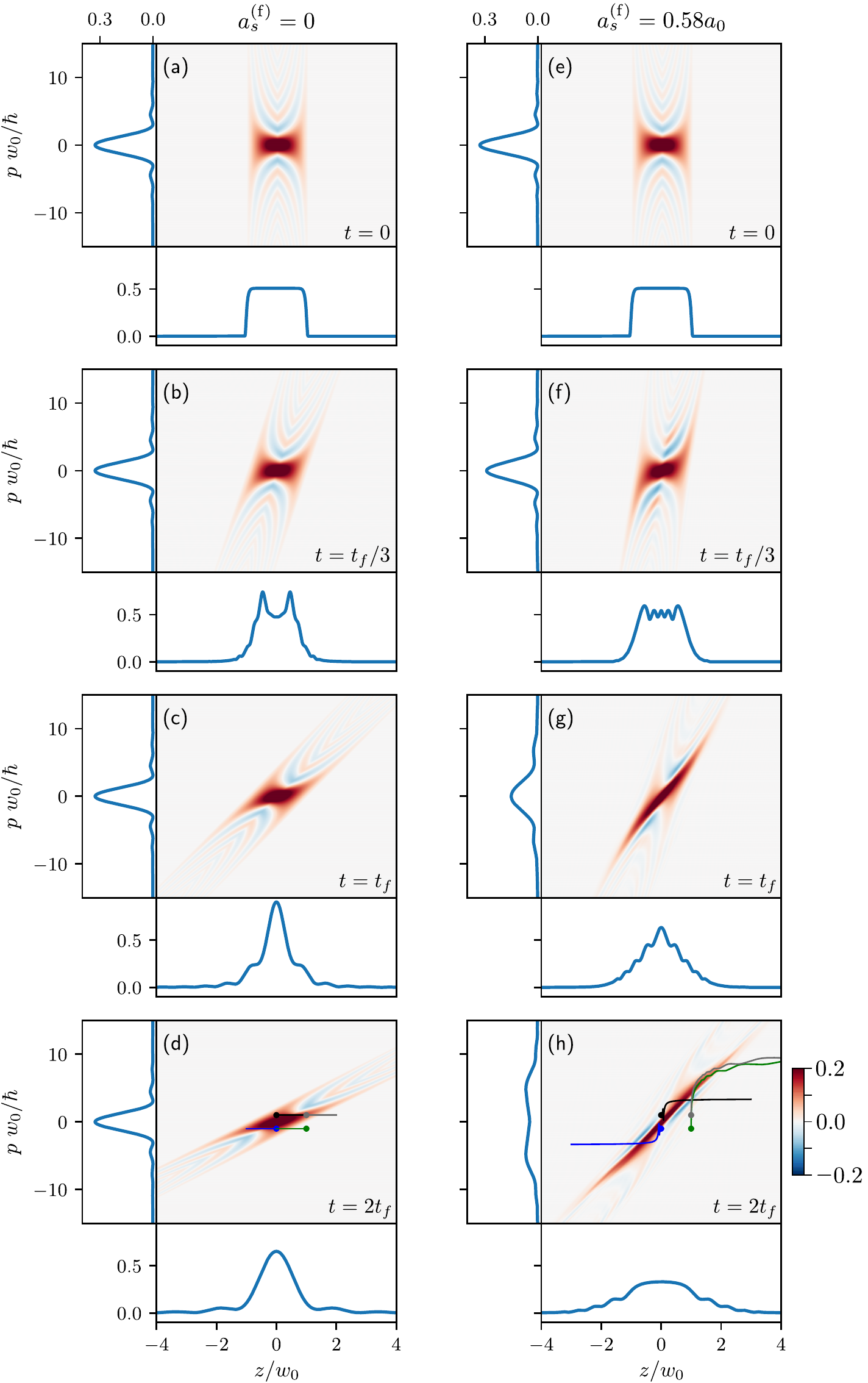}
		\caption{Diffractive focusing of a uniform BEC viewed from Wigner phase space. We illustrate the time evolution of the Wigner function corresponding to the ground state of the trapping potential $V_{\perp}+V_{10}$ after switching off only the longitudinal potential $V_{10}$, {\it and} changing the scattering length $a_s$ from its initial value $a_s=100\,a_0$ to its final one $a_s^{\rm (f)}=0$ (left column), or $a_s^{\rm (f)}=0.58\,a_0$ (right column). Here the red colors indicate large positive values of $W=W(z,p;t)$ and the blue ones mark domains of quantum phase space where $W$ assumes negative values as suggested by the color-code to the right of (h). The corresponding position and momentum distributions $|\varphi(z,t)|^2$ and $|\tilde\varphi(p,t)|^2$ are shown in the lower and left sub-figures, respectively. The classical trajectories $\{Z(t),P(t)\}$ governed by Eqs. \eqref{eq:classicaltrajectories}, \eqref{eq:classicaltrajectories2} and \eqref{eq:classicaltrajectories:hamiltonian} are displayed by different colors for different initial points in phase space. The focal time $t_f$ is a function of $a_s^{\rm (f)}$ as shown in Fig. \ref{fig_focusfactor_of_dynamical_interaction_diffl}.}
		\label{fig_phase_space_plots_2x4}
	\end{figure}

	Before we consider the case of interacting particles, we first recall \cite{goncalves_single-slit_2017, alonso_wigner_2011} the interaction-free dynamics ($a_s^{(\mathrm{f})}= 0$) of the Wigner function $W$, where the initial state is the ground state of the complete trapping potential $V=V_{\perp}+V_{10}$, as discussed in Section \ref{subsection-initial state}. In Figs. \ref{fig_phase_space_plots_2x4} (a)-(d) we display the Wigner functions for four different times, where the red and blue colors correspond to positive and negative values of $W = W(z,p;t)$, respectively. According to Eqs. \eqref{eq:wigner_position} and \eqref{eq:wigner_momentum}, the integration over the momentum or the position variable provides us with the position distribution $|\varphi(z,t)|^{2}$ (lower sub-figure), or the momentum distribution $|\tilde{\varphi}(p,t)|^{2}$ (left sub-figure).
	
	Figure \ref{fig_phase_space_plots_2x4} brings out most clearly the origin of the phenomenon of diffractive focusing. Indeed, at $t=0$, the Wigner function $W=W(z,p;0)$ exhibits both positive and negative values. During the free expansion, $t>0$, the parts of the Wigner function corresponding to $p>0$ ($p<0$) move to the right (left) along the straight lines $\left\{z+p t,p\right\}$, displayed in Fig. \ref{fig_phase_space_plots_2x4} (d) by different colors for four different initial points in phase space. These lines are parallel to the $z$-axis and describe the free classical motion, resulting from Eqs. \eqref{eq:classicaltrajectories}, \eqref{eq:classicaltrajectories2} and \eqref{eq:classicaltrajectories:hamiltonian} with $\tilde{g}=0$.
	
	At the time of focusing, $t = t_f$, the position distribution $|\varphi(z,t_f)|^{2}$ features a narrow maximum at $z = 0$. Indeed, integration over $p$ in Eq.~\eqref{eq:wigner_position} for fixed position $z$, yields a maximum only at the values of $z$, which correspond to the maximal values of $W=W(z,p,t_f)$, displayed by dark red color. According to Fig. \ref{fig_phase_space_plots_2x4} (a)-(d), this is the line $z = 0$ in phase space.
	In other words, focusing takes place at $z=0$, because at $t=t_f$ all negative parts of the initial Wigner function $W(z,p;0)$ have moved away from the $p$-axis \cite{case_diffractive_2012, goncalves_single-slit_2017}. However, they now subtract from the positive parts of the wings and make the distribution in space even narrower. 
	
	For $t>t_f$, the negative parts of $W(z,p,0)$ have moved further away from the line $z=0$. Since the positive parts of the original Wigner function are at lower momenta than the negative ones, they move slower and are therefore left at the center of the phase space. They are the origin of the spreading of the position distribution $|\varphi(z,t)|^{2}$.

	\subsection{Time evolution with atom-atom interaction}
	
	Next, we discuss the nonlinear time evolution of the Wigner function in the case of a non-vanishing atom-atom interaction, namely for the final value of the scattering length $a_s^{(\rm f)}=0.58\,a_0$. Figure \ref{fig_phase_space_plots_2x4} (e) presents the same initial Wigner function $W(z,p;0)$ as Fig. \ref{fig_phase_space_plots_2x4} (a).
	
	In contrast to the case of no atom-atom interaction, Figs. \ref{fig_phase_space_plots_2x4} (f)-(h) indicate that the parts of the Wigner function corresponding to $p>0$ ($p<0$) do not only move to the right (left) but also move up (down). This effect can be explained as follows. 
	
	For short times, $t\ll t_f$, we can neglect in Eq.~\eqref{1D GP effective} the kinetic energy term $(-\hbar^2/2m)\partial^2/\partial z^2$ compared to the interaction term $\tilde{g}|\varphi(z,t)|^{2}$ and arrive at the nonlinear equation
	\begin{equation}
		\label{eq_approx}
		i \hbar\frac{\partial}{\partial t}\varphi(z,t) \cong \tilde{g}|\varphi(z,t)|^2 \varphi(z,t),
	\end{equation}
	which is not easy to solve. 
	
	However, we note that Eq.~\eqref{eq_approx} conserves the quantity $|\varphi(z,t)|^{2}$, that is $\partial |\varphi(z,t)|^2/\partial t=0$, resulting in the simplified equation
	\begin{equation}
		i\hbar \frac{\partial}{\partial t}\varphi(z,t) = \tilde{g} |\varphi(z,0)|^2 \varphi(z,t)
	\end{equation}
	with the solution
	\begin{equation}
		\varphi(z,t) = \varphi(z,0)
		\exp\left(-\frac{i}{\hbar}\tilde{g}\,t|\varphi(z,0)|^{2}\right).
	\end{equation}
	
	Thus, for $t\ll t_f$, the wave function $\varphi = \varphi(z,t)$ picks up only a position-dependent phase determined by the initial distribution $|\varphi(z,0)|^2$ and the effective interaction strength $\tilde{g}$. The gradient $-\tilde{g}t \partial |\varphi(z,0)|^2 /\partial z $ of this phase defines the increase in momentum, which is a function of $z$.
	
	This result also follows from Eqs. \eqref{eq:classicaltrajectories} and \eqref{eq:classicaltrajectories:hamiltonian} and is displayed in Fig. \ref{fig_phase_space_plots_2x4} (h) by the classical trajectories corresponding to different initial points in phase space. Hence, due to this increase in momentum, the negative parts of the Wigner function get deformed and move faster away from the center compared to the case $\tilde{g} = 0$, resulting in the focus appearing at earlier times. This behavior is also confirmed by Fig. \ref{fig_free_time_evo_gdyn_100_densityplot} (b).
	
	For $t>t_f$, the position distribution $|\varphi(z,t)|^{2}$ spreads further, as shown in Fig. \ref{fig_phase_space_plots_2x4} (h), and reduces its amplitude. Thus, the interaction term $\tilde{g}|\varphi(z,t)|^{2}$ in Eq.~\eqref{1D GP effective} gets smaller compared to the kinetic energy term, and the evolution of the wave function $\varphi=\varphi(z,t)$ can be described solely by the Schr\"odinger equation. This effect is illustrated in Fig. \ref{fig_phase_space_plots_2x4} (h) by the classical trajectories, which in the long-time limit are again parallel to the $z$-axis.
	
	A closer look at the momentum distribution of Fig. \ref{fig_phase_space_plots_2x4} (h) reveals that two maxima are forming symmetrically around the origin of phase space due to the non-vanishing interaction. The momenta at which they occur directly depend on the final interaction strength $\tilde{g}^{\rm (f)}$ of the system and increases for larger $\tilde{g}^{\rm (f)}$. 
	For very long times the peaks smear out into a large momentum distribution at the center, however, if the interaction is set to zero beforehand, then the double-peak structure could be preserved. 
	
	In summary, we emphasize that diffractive focusing originates from the negative parts of the Wigner function \cite{case_diffractive_2012, goncalves_single-slit_2017}. According to the Hudson theorem \cite{Hudson}, a pure state with an initial Gaussian profile has a positive Wigner function at any point of the phase space, and therefore the state does not show diffractive focusing at all. A similar behavior occurs for any classical state. Hence, diffractive focusing for a given BEC gives us an opportunity to check whether this BEC is prepared in a non-Gaussian or non-classical state.
	
	\section{Conclusions and outlook}

	In this article we have studied the phenomenon of diffractive focusing of {\it interacting} matter waves employing analytical as well as numerical methods. We have proposed a straightforward implementation of this effect with an atomic BEC confined by a box-like trap as realized for instance in Ref.~\cite{gaunt_bose-einstein_2013}. The interaction of the atoms forming a BEC leads to the non-linearity in the GPE and is an essential ingredient in the preparation of a rectangular wave function, previously studied \cite{case_diffractive_2012, goncalves_single-slit_2017, weisman_diffractive_2017} in the context of Schr\"odinger waves, or the paraxial approximation in optics, and obtained by a rectangular slit.
	
	As benchmarks, we have identified the focusing factor and the focus time which both are functions of the strength of the atom-atom interaction. These measures allow us to derive the optimal conditions for observing this type of self-focusing of a BEC. Having identified the origin of diffractive focusing for interacting matter waves, illuminated by the time evolution of the Wigner function in phase space, we conclude that the cleanest realization occurs when the atom-atom interaction is switched off during the dynamics by a magnetic Feshbach resonance. 
	
	For the sake of simplicity we have restricted our treatment to a quasi-1D case. However, the effect of diffractive focusing takes also place for higher dimensions \cite{bialynicki-birula_-_2002} and could be realized with a 3D box potential \cite{gaunt_bose-einstein_2013} generated by blue-detuned laser light. Indeed, the focus factor achieves the value 4 for a cylindrically symmetric rectangular shape in {\it two} dimensions \cite{fresnel_diffractive_first}, whereas it is 1.8 for the rectangular initial profile in {\it one} dimension \cite{case_diffractive_2012, goncalves_single-slit_2017}. Moreover, diffractive focusing crucially depends on the initial profile, that is a more non-Gaussian or non-classical initial wavefunction results in stronger focusing. Thus, it would be useful to find the optimal initial profile giving rise to the best focusing. The problem of finding such an optimal state determined solely by the atom-atom interaction is highly non-trivial due to the non-linearity of the dynamical equation.

	We conclude by emphasizing that the results presented here can be immediately applied to other physical systems, whose dynamics is governed by the Gross-Pitaevskii-type equation, that is the cubic Schr\"odinger equation, for instance, to nonlinear optics \cite{agrawal_2012} and deep water surface water waves of moderate steepness \cite{mei_1983}.
	Moreover, the diffractive focusing can be used to generate bright sources of matter waves for dedicated applications in precision measurements \cite{bongs_taking_2019, Cronin_2009}. A more detailed discussion of these points goes beyond the scope of this article and has to be postponed to a future publication. 
	
	\section*{Acknowledgments}
	
	NG wishes to thank Eric Charron and Holger Ahlers for fruitful discussions during the early phase of this work and Timon Hilker  for helpful feedback about the manuscript.	This project is supported by the German Space Agency (DLR) with funds provided by the Federal Ministry for Economic Affairs and Energy (BMWi) under Grant Nos. 50WP1705 (BECCAL), 50WM1862 (CAL) and 50WM2060 (CARIOQA) as well as by the Deutsche Forschungsgemeinschaft (German Research Foundation) through CRC 1227 (Dq-mat) within Project No. A05.
	M.A.E. is thankful to the Center for Integrated Quantum Science and Technology ($\rm IQ^{ST}$) for its generous financial support. The research of the $\rm IQ^{ST}$ is financially supported by the Ministry of Science, Research and Arts, Baden-W\"urttemberg. W.P.S. is grateful to the Hagler Institute for Advanced Study at Texas A$\&$M University for a Faculty Fellowship, and to Texas A$\&$M AgriLife Research for the support of this work.

	\appendix

	\section{Dynamics of a Bose-Einstein condensate in a cigar-shaped trap}
	\label{Appendix_derivation_of_1D_GPE}

	We devote this appendix to the derivation of the effective 1D GPE describing the non-linear dynamics of a BEC along the longitudinal direction of a highly anisotropic cigar-shaped trapping potential. Here we approximate the complete wave function by the product of the transverse time-independent wave function of only the transverse coordinates $x$ and $y$, the time-dependent longitudinal wave function of the longitudinal coordinate $z$, as well as a time-dependent phase factor.
	This approach allows us to derive analytical formulas for the effective 1D interaction strength for (i) almost non-interacting and (ii) weakly interacting atoms.

	\subsection{Decoupling of transverse and longitudinal dynamics}
	
	To derive an equation governing the dynamics of a quasi-1D BEC which consists of $N$ atoms of mass $m$, we start from the 3D GPE \cite{pitaevskii_bose-einstein_2003}
	\begin{equation}
		\imag \hbar \pdiff[]{}{t} \psi\qty{\mathbf{r}, t} = \sqty{-\frac{\hbar^{2}}{2m} \pdiff[2]{}{\mathbf{r}} + V\qty{\mathbf{r}, t} + g N \abs{\psi\qty{\mathbf{r}, t}}^{2}}\psi\qty{\mathbf{r}, t}
		\label{eq_TB_GPE_full3D_Appendix}
	\end{equation}
	for the BEC wave function $\psi=\psi\qty{\mathbf{r}, t}$ with the external potential
	\begin{equation}
		V\qty{\mathbf{r}, t} \equiv V_{\perp}\qty{x,y} + V_{\mrm{Box}}\qty{z,t}
	\end{equation}
	being the sum of a harmonic trap 
	\begin{equation}
		V_{\perp}\qty{x,y} \equiv \frac{m}{2}\qty{\omega_{x}^{2} x^{2} + \omega_{y}^{2} y^{2}}
		\label{eq_transverspotential_Appendix}
	\end{equation}
	in the transverse directions determined by the trap frequencies $\omega_{x}$ and $\omega_{y}$, and  
	a box potential $V_{\mrm{Box}} = V_{\mrm{Box}}(z,t)$ yielding trapping in the $z$-direction. 
	
	In this article we consider a highly anisotropic cigar-shaped trapping geometry defined by the relation
	\begin{equation}
		L_{\perp} \ll L_z ,
		\label{eq_Lperp_ll_Lz_Appendix}
	\end{equation}
	where $L_z$ is the longitudinal characteristic length of the external potential and $L_{\perp} \equiv \sqrt{L_{x} L_{y}}$ is the transverse one with $L_{x}\equiv \sqrt{\hbar/m\omega_{x}}$ and $L_{y}\equiv \sqrt{\hbar/m\omega_{y}}$.
	
	In this case, as shown in \ref{sec:App_B_chemPot_energy}, the total energy per particle of the BEC is approximately given by the relation
	\begin{equation}
		\label{Energy_appendix}
		\frac{E}{N} \cong \hbar\omega_\perp \left(\frac{N a_s}{L_z}\right)^{1/2}
	\end{equation}
	with $\omega_{\perp}\equiv \sqrt{\omega_x \omega_y}$.
	
	Hence, for 
	\begin{equation}
		0 \leq N a_s \ll L_z,
		\label{eq_Nas_ll_Lz_Appendix}
	\end{equation}
	the total energy per particle $E/N$ is much smaller than the characteristic energy scale $\hbar\omega_\perp$ of the transverse direction, making it  impossible to drive collective excitations in that direction as long as the energy of the system is conserved. Indeed, we effectively freeze out the transverse dynamics~\cite{olshanii_atomic_1998, salasnich_effective_2002, mateo_effective_2008} 
	
	Consequently, the total wave function
	\begin{equation}
		\psi\qty{\mathbf{r}, t} \equiv \Phi_0 \left(x,y\right)
		\varphi\qty{z,t}\exp\left(-\frac{\imag}{\hbar}\varepsilon_0 t\right)
		\label{eq_TB_seperation_condition_Appendix}
	\end{equation}
	can be approximated by the product of the real-valued wave function $\Phi_0 = \Phi_0 \left(x,y\right)$ describing the ground state in the transverse direction, the wave function $\varphi = \varphi\qty{z,t}$ along the $z$-direction, and a time-dependent phase factor, where the constant $\varepsilon_0$ shall be determined later as to simplify the equations. 
	
	Moreover, the function $\Phi_0$ is chosen to be normalized, that is
	\begin{equation}
		\label{eq:Phi0_normalization_Appendix}
		\int \dd x \dd y \,\Phi_0^2=1.
	\end{equation}
	
	When we insert our ansatz, Eq.~\eqref{eq_TB_seperation_condition_Appendix}, into the 3D GPE, Eq.~\eqref{eq_TB_GPE_full3D_Appendix}, we obtain the identity
	\begin{eqnarray}
		\label{eq:full_equation_Phi0_varphi_Appendix}
		\Phi_0\left(i\hbar\frac{\partial\varphi}{\partial t}\right)+
		\varepsilon_0 \Phi_0\varphi &=&
		\left[-\frac{\hbar^2}{2m}\left(\frac{\partial^2 \Phi_0}{\partial x^2}+\frac{\partial^2\Phi_0}{\partial y^2}\right)+V_\perp \Phi_0\right]\varphi \nonumber \\
		&+&\Phi_0\left(-\frac{\hbar^2}{2m}\frac{\partial^2\varphi}{\partial z^2}+V_{\rm Box}\varphi\right)
		+gN|\varphi|^2\Phi_0^2\Phi_0\varphi .
	\end{eqnarray}
	
	Finally, we multiply both sides of Eq.~\eqref{eq:full_equation_Phi0_varphi_Appendix} from the left by $\Phi_0$, integrate over $x$ and $y$, and arrive at the non-linear equation
	\begin{equation}
		\label{phi_equation_Appendix}
		\imag \hbar \pdiff[]{}{t} \varphi= \left(-\frac{\hbar^{2}}{2m} \pdiff[2]{}{z} + V_{\mrm{Box}}(z)+\tilde{g}|\varphi|^2\right)\varphi
	\end{equation}
	for the longitudinal wave function $\varphi=\varphi(z,t)$, with the effective interaction strength
	\begin{equation}
		\tilde{g} \equiv g N c_{\perp}
		\label{dimension_g_Appendix}
	\end{equation}
	determined by the interaction constant $g$, Eq.~\eqref{g3D}, the number of particles $N$, and the integral 
	\begin{equation}
		c_{\perp} \equiv \int \dd x \dd y \, \Phi_0^4.
		\label{c_perp-def_Appendix}
	\end{equation}
	Here we have made use of Eq.~\eqref{eq:Phi0_normalization_Appendix} and have chosen the constant
	\begin{equation}
		\label{E0-def_Appendix}
		\varepsilon_0 \equiv \int \dd x \dd y \, \Phi_0 \sqty{-\frac{\hbar^{2}}{2m}\left(\pdiff[2]{}{x}+\pdiff[2]{}{y}\right) + V_{\perp}\qty{x,y} }\Phi_0
	\end{equation}
	to simplify Eq.~\eqref{phi_equation_Appendix}.
	
	As a result, we have derived the 1D GPE \eqref{phi_equation_Appendix} which describes the longitudinal dynamics of a quasi-1D BEC characterized by the two inequalities Eqs. \eqref{eq_Lperp_ll_Lz_Appendix} and \eqref{eq_Nas_ll_Lz_Appendix}. In order to employ Eq.~\eqref{phi_equation_Appendix}, we first have to find the ground-state wave function $\Phi_0$ of the transverse direction and then evaluate the effective interaction strength $\tilde{g}$, Eq. \eqref{dimension_g_Appendix}.

	According to Eqs. \eqref{eq_Lperp_ll_Lz_Appendix} and \eqref{eq_Nas_ll_Lz_Appendix}, there exist two distinct cases where both $\Phi_0$ and $\tilde{g}$ can be calculated analytically, namely the limit of almost non-interacting atoms, $0 \leq N a_s \ll L_\perp \ll L_z$, and the case of weakly-interacting atoms, $L_\perp \ll N a_s \ll L_z$. In the next sections we consider these two situations.

	\subsection{Almost non-interacting atoms}
	\label{Appendix_calc_gauss}
	
	For almost non-interacting atoms with $0 \leq N a_s \ll L_\perp \ll L_z$, we neglect the interaction term $g N |\psi|^{2}$ in the 3D GPE \eqref{eq_TB_GPE_full3D_Appendix}, and the equation becomes approximately separable in the coordinates $x$, $y$, and $z$. As a result, the transverse wave function $\Phi_0$ in the ansatz, Eq.~\eqref{eq_TB_seperation_condition_Appendix}, for $\psi$ coincides with the wave function
	\begin{equation}
		\label{Phi0 - small g}
		\Phi_0^{(\rm ho)}(x,y)\equiv \frac{1}{\sqrt{\pi L_x L_y}}\exp\left(-\frac {x^2}{2L_x^2}-\frac {y^2}{2L_y^2}\right)
	\end{equation}
	of the ground state of a 2D harmonic oscillator.
	
	By inserting Eq.~\eqref{Phi0 - small g} into the definitions for $c_{\perp}$, Eq.~\eqref{c_perp-def_Appendix}, and $\varepsilon_0$, Eq.~\eqref{E0-def_Appendix}, and performing the integration over $x$ and $y$, we obtain the explicit expressions 
	\begin{equation}
		c_{\perp}= \frac{1}{2\pi L_x L_y} \equiv
		\frac{1}{2\pi L_{\perp}^2}
		\label{appendix1_cperp}
	\end{equation}
	for the parameter $c_{\perp}$ and 
	\begin{equation}
		\label{appendix_E0}
		\varepsilon_0=\frac{1}{2}\hbar(\omega_x+\omega_y)
	\end{equation}
	for the constant $\varepsilon_0$.
	
	This approach to quasi-1D BECs has already been discussed in similar ways by other groups~\cite{olshanii_atomic_1998, salasnich_effective_2002, mateo_effective_2008}. Our results exactly coincide with their findings for the same order of approximation.

	\subsection{Weakly interacting atoms}
	\label{Appendix_calc_TF}
	
	In the case of weakly interacting atoms, that is for $L_\perp \ll N a_s \ll L_z$, the interaction term $g N |\psi\qty{\mathbf{r}, t}|^{2}$ in the 3D GPE \eqref{eq_TB_GPE_full3D_Appendix} is the leading one and we can apply the Thomas-Fermi approximation \cite{pitaevskii_bose-einstein_2003} by neglecting the kinetic term when determining the ground-state wave function. Starting from the stationary solution
	\begin{equation}
		\psi\qty{\mathbf{r}, t}=\phi(x,y,z)\exp\left(-\frac{\imag}{\hbar} \mu t\right),
	\end{equation}
	we thus obtain the ground-state wave function
	\begin{equation}
		\label{eq_thomas_fermi_equaton}
		\phi(x,y,z)  = \sqrt{\frac{\mu - V(x,y,z)}{gN}}\,
		\Theta\left[\mu - V(x,y,z)\right].
	\end{equation}
	Here $\Theta$ denotes the Heaviside function and
	\begin{equation}
		\label{mu-TF-result}
		\mu = \qty{\frac{mgN \omega_{x} \omega_{y}}{2\pi L_z}}^{\frac{1}{2}}
	\end{equation}
	is the chemical potential derived in \ref{sec:App_B_chemPot_energy} when the box potential is approximated by infinitely high potential walls separated by $2L_z$.
	
	As a result, within the Thomas-Fermi approximation, the total wave function $\phi=\phi(x,y,z)$ given by Eq.~\eqref{eq_thomas_fermi_equaton} is again the product 
	\begin{equation}
		\phi(x,y,z)=\Phi_0(x,y)\varphi_0(z)
	\end{equation}
	of the normalized transverse wave function
	\begin{equation}
		\label{Phi0-TF-result}
		\Phi_0\qty{x,y} \equiv \left[2 L_{z}\frac{\mu - V_{\perp}(x, y)}{g N}
		\right]^{\frac{1}{2}}\Theta\left[\mu - V_{\perp}(x, y)\right] 
	\end{equation}
	and the longitudinal wave function 
	\begin{equation}
		\varphi_0\qty{z} \equiv \frac{1}{\sqrt{2 L_{z}}}\Theta(L_{z} - \abs{z}),
	\end{equation}
	with $V_{\perp} = V_{\perp}(x,y)$ and $\mu$ given by Eqs. \eqref{eq_transverspotential_Appendix} and \eqref{mu-TF-result}, respectively.  
	
	By inserting Eq.~\eqref{Phi0-TF-result} into Eq.~\eqref{c_perp-def_Appendix}, we obtain the explicit expression 
	\begin{equation}
		c_\perp \cong \frac{1}{2\pi L_{\perp}^2} \left(\frac{8}{9}\frac{L_z}{Na_s}\right)^{\frac{1}{2}}
		\label{apendix_2_c_perp}
	\end{equation}
	for the parameter $c_{\perp}$.
	
	Moreover, inserting Eq.~\eqref{Phi0-TF-result} into Eq.~\eqref{E0-def_Appendix}, and neglecting the second-order derivatives over $x$ and $y$, we arrive at the formula
	\begin{equation}
		\label{appendix_E0_TF}
		\varepsilon_0\cong \int \dd x \dd y \, V_{\perp}\qty{x,y}\Phi_0^2=\frac{1}{3}\mu
	\end{equation}
	for the constant $\varepsilon_0$, where $\mu$ is given by Eq.~\eqref{mu-TF-result}.
	
	We emphasize that the expression, Eq. \eqref{apendix_2_c_perp}, for $c_\perp$ is still obtained in the regime where the motion along the transverse direction is effectively frozen out ($N a_s \ll L_z$), but in contrast to the previous case the interaction between the particles is taken into account when determining the shape of the transverse ground-state wave function $\Phi_0$. The comparison between the solutions of the effective 1D GPE \eqref{phi_equation_Appendix} with $c_{\perp}$ given by Eq. \eqref{apendix_2_c_perp}, and the 3D GPE \eqref{eq_TB_GPE_full3D_Appendix}, presented in Fig. \ref{fig_comparison3D}, shows that, for the parameters considered in this article, our expression, Eq. \eqref{apendix_2_c_perp}, for $c_\perp$ describes the dynamics more accurately than the standard formula, Eq. \eqref{appendix1_cperp}, corresponding to weakly interacting atoms.
	
	We conclude this discussion by noting that the case of even stronger atom-atom interaction, when $L_z \ll N a_s$, can be treated in a similar way. Indeed, the dynamics along the transverse direction is then much faster compared to the longitudinal direction due to the relation $L_\perp \ll L_z$. Here one can perform the adiabatic approximation~\cite{salasnich_effective_2002, mateo_effective_2008} to factorize the total wave function and to describe the longitudinal dynamics of the quasi-1D BEC.

	\section{Thomas-Fermi approximation: chemical potential and energy of a Bose-Einstein condensate}
	\label{sec:App_B_chemPot_energy}
	
	The decoupling of the longitudinal and transverse degrees of freedom analyzed in \ref{Appendix_derivation_of_1D_GPE}
	rests on the estimate, Eq. \eqref{Energy_appendix}, of the total energy of the BEC per particle in terms of the characteristic energy of the transverse motion. In this appendix we use the Thomas-Fermi approximation and derive this estimate by first obtaining the analytical expression for the chemical potential of a BEC governed by the 3D GPE~\eqref{eq_TB_GPE_full3D_Appendix}. By elementary integration of the relation between the chemical potential and the energy, we then arrive at the desired estimate.

	\subsection{Chemical potential}
	
	The chemical potential $\mu$ of a BEC within the Thomas-Fermi approximation follows from the normalization condition
	\begin{equation}
		\label{TF-normalization condition}
		I = \int \dd x \dd y \dd z \abs{\phi\qty{x,y,z}}^{2} = 1
	\end{equation}
	of the Thomas-Fermi wave function \cite{pitaevskii_bose-einstein_2003}
	\begin{equation}
		\label{eq_thomas_fermi_equaton_appendix_B}
		\phi(x,y,z)  = \sqrt{\frac{\mu - V(x,y,z)}{gN}}\,
		\Theta\left[\mu - V(x,y,z)\right].
	\end{equation}
	Here $\Theta$ is the Heaviside function. 
	
	In order to derive an analytical expression for $\mu$, we approximate $V_{\rm Box}$ by the potential of infinitely high  walls    
	\begin{equation}
		\label{V(z) longitudinal}
		V_{\rm Box} (z) \cong \left\{
		\begin{array}{ll}
			0, &  |z|\leq L_z \\ \infty, & |z|> L_z
		\end{array}
		\right.
	\end{equation}
	separated by $2L_z$.
	
	According to Eq.~\eqref{eq_thomas_fermi_equaton_appendix_B}, only the points $\{x,y,z\}$ obeying the inequality $V(x,y,z)\leq \mu$ contribute to the integral in Eq.~\eqref{TF-normalization condition}. Hence, the regions of integration in Eq.~\eqref{TF-normalization condition} are given by
	\begin{equation}
		\frac{x^2}{b_x^2}+\frac{y^2}{b_y^2}\leq 1\;\;\;{\rm and}\;\;\;
		-L_z\leq z\leq L_z,
	\end{equation}
	where $b_x^2\equiv 2\mu/(m\omega_x^2)$ and $b_y^2\equiv 2\mu/(m\omega_y^2)$.
	
	By introducing the polar coordinates $x\equiv b_x r\cos\theta$ and $y\equiv b_y r\sin\theta$ with $0\leq r \leq 1$ and $0\leq \theta \leq 2\pi$, we arrive at 
	\begin{eqnarray}
		I = &\frac{b_x b_y}{gN}\int_{-L_{z}}^{L_z}\dd z \int_{0}^{1} r\dd r \int_{0}^{2\pi}  \dd \theta \left(\mu-\mu r^{2}\right),
	\end{eqnarray}
	where we have used the identity $V_{\perp}(b_x r\cos\theta,b_y r \sin\theta)=\mu r^2$, which then leads us to 
	\begin{equation}
		I = \pi\frac{\mu b_x b_y L_z}{gN}.
	\end{equation}
	
	With the definitions of $b_x$ and $b_y$ together with the normalization condition, Eq.~\eqref{TF-normalization condition}, we find the explicit expression 
	\begin{equation}
		\mu = \qty{\frac{mgN \omega_{x} \omega_{y}}{2\pi L_z}}^{\frac{1}{2}}
		\label{mu-TF-result_appendix_B}
	\end{equation}
	for the chemical potential of a BEC being confined by an infinitely high box potential along the $z$-axis and two harmonic potentials along the $x$- and $y$-direction. We emphasize that Eq.~\eqref{mu-TF-result_appendix_B} is valid for arbitrary length scales $L_\perp$ and $L_z$ of the external potentials.

	\subsection{Energy}
	A similar calculation can be performed to find the total energy~\cite{pitaevskii_bose-einstein_2003}
	\begin{equation}
		E = N \int \dd x \dd y \dd z \left[ V(x,y,z) \abs{\phi\qty{x,y,z}}^{2} + \frac{g}{2}\abs{\phi\qty{x,y,z}}^{4} \right]  
	\end{equation}
	of a BEC within the Thomas-Fermi approximation. 
	
	However, the more convenient approach consists of inserting the result, Eq.~\eqref{mu-TF-result_appendix_B}, for the chemical potential into the definition~\cite{pitaevskii_bose-einstein_2003}
	\begin{equation}
		\mu = \frac{\dd E}{\dd N},
		\label{eq_defnition_mu_E_appendix_B}
	\end{equation}
	which we can directly integrate to obtain 
	\begin{equation}
		E = \frac{2}{3} N \mu .
		\label{eq_result_E_mu_appedix_B}
	\end{equation}
	
	This relation \cite{Meister2017} coincides with the one for a purely harmonically trapped BEC in two dimensions. 
	
	By inserting Eqs.~\eqref{mu-TF-result_appendix_B} and \eqref{g3D} for the interaction constant $g$ into Eq.~\eqref{eq_result_E_mu_appedix_B}, we arrive at the expression
	\begin{equation}
		\frac{E}{N} = \frac{2\sqrt{2}}{3} \hbar\omega_\perp \left(\frac{N a_s}{L_z}\right)^{1/2}.
	\end{equation}
	
	With $2\sqrt{2}/3\approx 0.943$, this yields the estimate
	\begin{equation}
		\frac{E}{N} \cong \hbar\omega_\perp \left(\frac{N a_s}{L_z}\right)^{1/2}
	\end{equation}
	for the total energy per particle within the Thomas-Fermi approximation.

	\section{Thomas-Fermi wave function for optical trapping potential}
	
	\label{fidelity_rect}
	In this appendix we derive the wave function for ground state of the potential $V_l$, Eq. \eqref{eq_pep_of_initial_state__approximatepotential}, within the Thomas-Fermi approximation. Indeed, the Thomas-Fermi profile of the stationary solution 
	
	\begin{equation}
		\varphi(z,t)=\varphi_{TF}(z)\exp\left(-\frac{i}{\hbar}\mu_{TF}t\,\right)
	\end{equation}
	of the 1D GPE 
	\begin{equation}
		\label{1D GP effective_appendix}
		\imag \hbar \pdiff[]{}{t} \varphi\qty{z,t} = \left(-\frac{\hbar^{2}}{2m} \pdiff[2]{}{z} + V_{\mrm{Box}}(z) + \tilde{g}|\varphi|^2\right)\varphi\qty{z,t},
	\end{equation}
	reads \cite{pitaevskii_bose-einstein_2003}
	\begin{equation}
		\label{phi-TF}
		\varphi_{TF}(z)=\left[\frac{\mu_{TF} - V_{l}(z)}{\tilde{g}}
		\right]^{\frac{1}{2}}\Theta\left[\mu_{TF} - V_{l}(z)\right].
	\end{equation}
	Here we have used the potential $V_l = V_l(z)$, Eq.~\eqref{eq_pep_of_initial_state__approximatepotential}, for $V_{\rm Box}$ in Eq.~\eqref{1D GP effective_appendix}. 
	
	The chemical potential $\mu_{TF}$ is determined by the normalization condition
	\begin{equation}
		\label{phi-TF-normalization}
		\int_{-\infty}^{\infty}dz|\varphi_{TF}(z)|^2=1.
	\end{equation}
	
	According to Eq.~\eqref{eq_pep_of_initial_state__approximatepotential} the LG potential $V_l$ can be approximated by
	\begin{equation}
		\label{V_L_Appendix2}
		V_l(z)\approx v_l\left(\frac{z}{w_0}\right)^{2l},
	\end{equation}
	where we have introduced the abbreviation
	\begin{equation}
		\label{v_l}
		v_l\equiv \frac{2^l}{4\pi l!}\frac{\hbar \Gamma^2}{\Delta }\frac{P}{I_s w_{0}^2}.
	\end{equation}
	
	The Thomas-Fermi distance $z_{TF}$ defined by the condition $\mu_{\mathrm{TF}}= V_l(z_{\mathrm{TF}})$ follows from the potential $V_l$ given by Eq.~\eqref{V_L_Appendix2} as 
	\begin{equation}
		\label{mu-TF-z-TF-1}
		z_{TF}=w_0\left(\frac{\mu_{TF}}{v_l}\right)^{\frac{1}{2l}}.
	\end{equation}
	
	Hence, the normalization condition, Eq. \eqref{phi-TF-normalization}, of the Thomas-Fermi wave function, Eq.~\eqref{phi-TF}, takes the form
	\begin{equation}
		\label{normalization_AppendixB_calc}
		1 = \frac{\mu_{TF}}{\tilde{g}}\int_{-z_{TF}}^{z_{TF}}dz
		\left(1-\frac{z^{2l}}{z_{TF}^{2l}}\right)=\frac{4l}{2l+1}\frac{\mu_{TF}}{\tilde{g}}z_{TF}.
	\end{equation}

	When we combine Eqs. \eqref{mu-TF-z-TF-1} and \eqref{normalization_AppendixB_calc}, we obtain the expression
	\begin{equation}
		\label{mu-TF-result_dependence_l}
		\mu_{TF}=\left(1+\frac{1}{2l}\right)^{\frac{2l}{2l+1}}\left(v_l\frac{\tilde{g}^{2l}}{2^{2l}w_0^{2l}}\right)^{\frac{1}{2l+1}}
	\end{equation}
	for the chemical potential in terms of the order $l$ and the parameters $w_0$ and $v_l$ of the LG mode, which is now used to calculate the Thomas-Fermi wave function $\varphi_{TF}$, Eq. \eqref{phi-TF}.

	
	

	\section*{References}
 	\bibliographystyle{nature}
	\bibliography{bibliographymaster.bib}

\end{document}